# Salinity-Dependent Interfacial Phenomena Towards Hydrovoltaic Device Optimization

Tarique Anwar and Giulia Tagliabue*

*Laboratory of Nanoscience for Energy Technologies (LNET), STI, École Polytechnique Fédérale de Lausanne (EPFL), Lausanne 1015, Switzerland*

*Email: giulia.tagliabue@epfl.ch

## Abstract

Evaporation-driven fluid flow in porous or nanostructured materials has recently opened a new paradigm for renewable energy generation by converting thermal energy into electrical energy via an electrokinetic pathway. Despite recent progress, major fundamental questions remain regarding the interfacial phenomena governing these so-called hydrovoltaic (HV) devices. Together with the lack of modelling tools, this limits the performance and application range of this emerging technology. By leveraging ordered arrays of Silicon nanopillars (NP) and developing a quantitative multiphysics model to study their HV response across a wide parameter space, this work reveals the complex interplay of surface-charge, liquid properties, and geometrical parameters, including previously unexplored electrokinetic interactions. Notably, we find that ion-concentration-dependent surface charge, together with ion mobility, dictates multiple local maxima in open circuit voltage, with optimal conditions deviating from conventional low-concentration expectations. Additionally, assessing the HV response up to molar concentrations, we provide unique evidence of ion adsorption and charge inversion for a number of monovalent cations. This effect interestingly enables the operation of HV devices even at such high concentrations. Finally, we highlight that, beyond electrokinetic parameters, geometrical asymmetries in the device structure generate an electrostatic potential that augments HV performance. Overall, our work, which lies in between single nanochannel studies and macro-scale porous system characterization, demonstrates that evaporation-driven HV devices can operate across a wide range of salinities, with optimal operating conditions being dictated by distinct interfacial phenomena. Thus it offers crucial insight and a design tool for enhancing the performance of evaporation-driven HV devices and paves the way to their broader applicability across the salinity scale of natural and processed waters.

## Introduction

Water evaporation is ubiquitous in nature, occurring spontaneously even without solar illumination. It enables continuous energy exchange through the water cycle and is a currently untapped renewable energy source [1–4]. As an example, the total power generation potential of natural evaporation from lakes and reservoirs in the contiguous United States was estimated at 325 gigawatts, > 69% of the US electric energy production rate in 2015[3]. Various evaporation-driven devices have been explored which could be broadly classified as tandem devices[5,6], hybrid systems[7,8], and self-powered generators[9–13]. In particular, in 2017, among the self-powered generators, it was demonstrated that evaporation-driven flow through a functionalized porous carbon film could reliably generate sustained voltages up to 1 V and 100 nA current at ambient conditions[11]. Following this, various approaches such as a decrease of fluidic impedance, electrical resistance[14], and surface functionalization[15] of the material structure have been demonstrated to improve the electrical power from 100 nW to ∼ 10 µW from a cm-sized device. More recently, the use of Silicon has been shown to improve the current density significantly via ion-electron coupling through coulombic interaction, resulting in the enhancement in power density by two orders of magnitude[12,16–18]. Overall, these results have attracted growing interest in the field of evaporation-driven HV devices[1,12,13,19–21]. However, due to the complex micro- and nanostructure of the devices investigated so far, there remains a lack of fundamental understanding related to the underlying interfacial phenomena. Furthermore, the concentration-dependent performance of HV systems remains largely unknown. Interestingly, though, a recent study has shown that exploiting the synergistic effect of a nanofluidic diode and redox reactions[22] a substantial amount of energy can be harvested in high salinity conditions. Thus, it would be extremely appealing to explore the possibility of designing HV devices that can be operated optimally across salinity scales of natural water (fresh and seawater) and processed water (brine).

On the other hand, pressure-driven flow in functionalized nanochannels has been widely studied for electrical energy conversion via the streaming current pathway[23–26]. In particular, ion transport in nanopores with a diameter spanning from a few microns to sub-nanometers has been carefully studied in the Iontronic community[27–31]. This has shown that surface charge, pore geometry, ionic mobility, etc., are critical for the device output. In contrast, almost all the previous reports on HV used bulk porous material[6,9,11,12], which limits the understanding of the role of geometrical parameters of the nanostructured material on energy conversion. Additionally, previous studies on HV devices focused on a narrow parameter space, which could not highlight the relevance of many interesting phenomena observed in other electrokinetic devices, such as regime of EDL overlap, concentration dependent surface charge[25,27], non-linear electrostatic screening[32], effect of ionic mobility, and charge inversion[33,34]. Therefore, there remains a need to understand the fundamental role of a variety of electrokinetic phenomena in evaporation-driven HV devices. In particular, the interplay of structural, interfacial, and fluidic properties must be clarified in order to identify opportunities and guidelines for future device design and performance optimization.

Here we present a controlled study of evaporation-driven HV devices that reveals distinct interfacial phenomena across salinity scales and quantifies their impact on performance. By using a nanostructuring approach applied to arrays of Silicon nanopillars (Si NPs), we systematically change the solid-liquid interfacial area, the liquid confinement size as well as the ion concentration and type. We then correlate their effect to the open circuit voltage ($V_{OC}$) and power output ($P_{max}$) of the device. Importantly, by developing a quantitative Multiphysics model to interpret the experimental results, we provide deeper insights into the dominant solid-liquid interfacial phenomena at different regimes. For the first time, we show that chemical equilibrium at the interface plays a critical role in all tested conditions. In fact, it controls the dissociation of surface groups, leading to an electrolyte concentration dependent surface charge[35]. Additionally, we highlight that intrinsic geometrical asymmetries in the device structure boost the voltage output independently from electrokinetic modulation. In terms of concentration-dependent performance, we confirm that, at low concentrations (< 1 mM), geometry must be optimized to exploit electrical double layer overlap. Most interestingly, at intermediate concentrations (< 0.1 M), we demonstrate that, due to the interplay of concentration dependent surface-charge and ion mobility, multiple $V_{OC}$ local maxima exist, allowing for optimal operating conditions that deviate from low-concentration expectations. Additionally, by assessing the HV device response at high concentrations (up to 4M), we uniquely show evidence of ion adsorption and charge inversion for several monovalent cations ($Li^+$ and $K^+$). We also report viable power levels under these extreme operating conditions. Overall, our controlled nanostructuring approach, which lies in between nanofluidic studies and macro-scale porous system characterization, highlights the importance of controlling electrokinetic and geometrical parameters in HV systems. Our results also show that, by leveraging distinct interfacial phenomena, evaporation-driven HV devices can be developed for a broader range of concentrations, and hence applications, than previously thought. Therefore, in combination with the presented predictive model, this work opens new avenues for the engineering of evaporation-driven HV devices with improved performance and expanded applicability across the entire salinity scale of natural and processed water, from freshwater to brine.

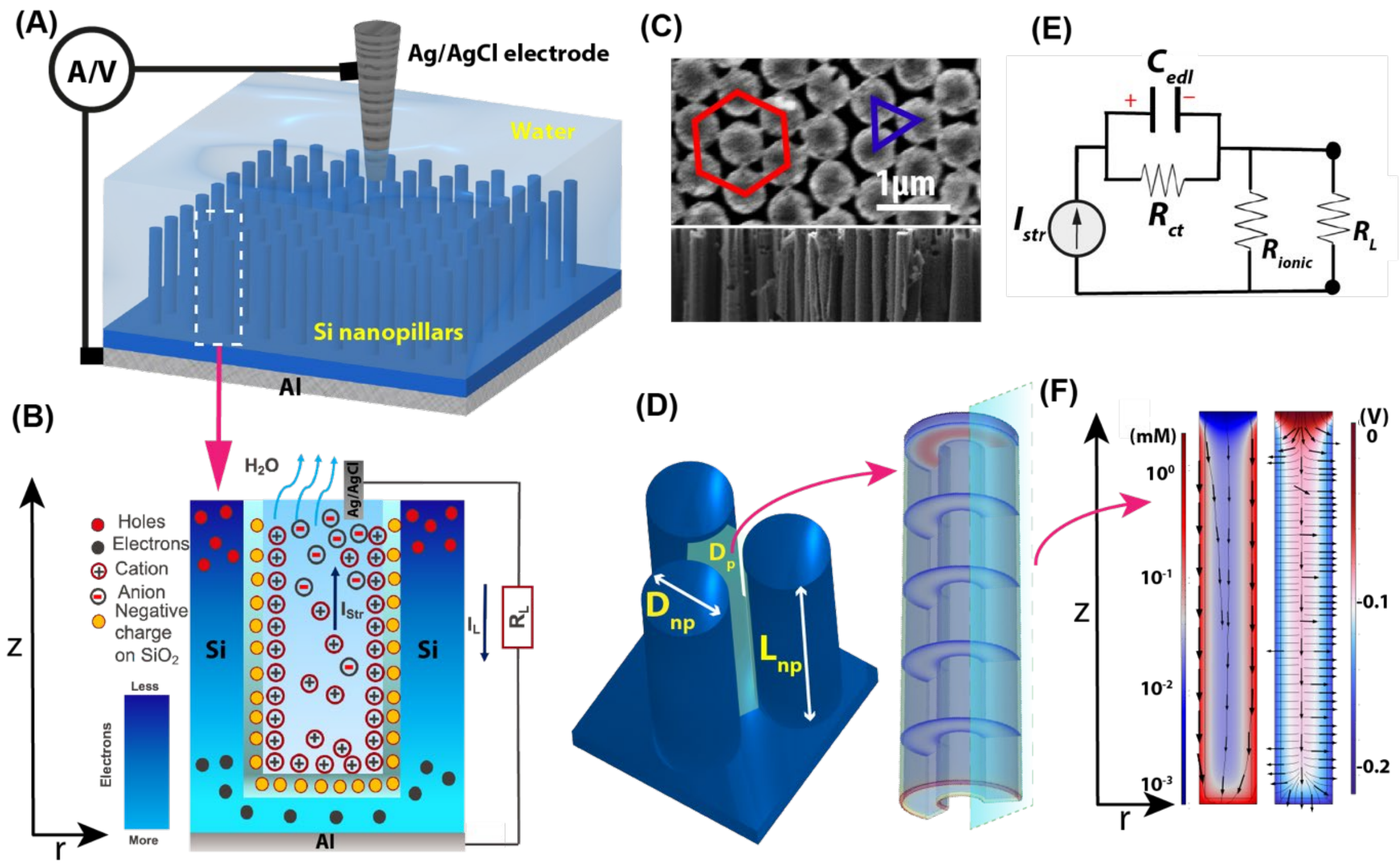

**Figure 1: Mechanism of the HV device made of Silicon NPs array. A)** Schematic of the electrical measurement setup for Silicon NPs hydrovoltaic device using a single Ag/AgCl electrode. **B)** The cross-section of the nanoconfinement formed due to the separation between adjacent NPs. The surface contains a negative surface charge (yellow circles) that forms an EDL. The evaporation-driven flow causes the streaming of ions, $I_{str}$, and the inherent asymmetry causes a difference in EDL potential between the top and bottom surface even when $I_{str} = 0$. **C)** (Top) SEM image showing a top view of the fabricated array of silicon NPs. The red line shows the hexagonal arrangement of the NPs, while the blue line represents the triangular unit cell used for simulation. (Bottom) Cross-sectional view. **D)** (Left) 3D schematic of the triangular unit cell of the hexagonally arranged NPs showing the geometrical parameters of the nanostructures (pillar diameter, $D_{np}$, pillar length, $L_{np}$ and mean pore diameter, $D_p$). (Right) Annular cylindrical nanopore geometry was used for simulations, including the calculated electrical potential distribution (see also panel (F)). **E)** Equivalent electrical circuit in which streaming of ions and the asymmetrical EDL potential between top and bottom are represented by a current source, and a capacitor in parallel with a resistor respectively. The ionic resistance and external load are represented by the appropriate resistances. **F)** Vertical cut-plane of the simulated cylindrical nanopore in panel (D) showing the counter-ion concentration distribution, with ion flux (left), and the electrical potential distribution, with electric field lines (right). The bulk ionic concentration is 10 μm KCl for the given simulation results.

## Results and discussion

Understanding the fundamental mechanisms of voltage and current generation in evaporation-driven HV devices requires control over the solid-liquid interface properties and the liquid nanoconfinement, i.e., nanochannel geometry. Our devices consist of cm-scale regular arrays of Si NPs etched in a p-type Si wafer (Figure 1A). Using a combination of colloidal lithography and metal-assisted chemical etching (see Methods), we fix the pitch (p = 600 nm) of the hexagonal array of NPs (see red line in Figure 1C) while varying their diameter ($D_{np}$) and length ($L_{np}$) in the range 420 nm – 560 nm and 1.23 µm – 4.4 µm respectively. The corresponding mean pore diameter $D_p$ ranges between 140 nm and 280 nm (Figure 1D and Figure S1). Therefore, changing the Si NPs dimensions directly controls the nanochannel geometry and the solid-liquid surface area ($A_{S-L}$), as needed.

For hydrovoltaic testing, the sample is placed inside an HV cell and then wetted with 150 µl of deionized water containing different salts (KCl, LiCl, CsCl) of varying concentrations (from 1 µM to 4 M) while being placed in ambient conditions (T = 22-24°C; and humidity = 25-30%). Thus, evaporation occurs naturally (Figure S2). The electrical response is measured using an Ag/AgCl electrode, placed in the liquid right above the Si NPs, and an aluminum contact, previously deposited on the back surface of the Si wafer (Figure 1A and Methods). During the electrical measurements, the Ag/AgCl electrode and silicon substrate do not contact each other, and the electrical circuit is complete as soon as the liquid is dispensed, thereafter voltage and current can be measured. Figure 1B shows a zoom-in view of a single nanochannel within our device. Upon wetting, the native silicon oxide layer on the surface of the Si NPs[36] will dissociate, resulting in a net negative surface charge σ (yellow circles), for pH above the isoelectric point.[35] Concurrently, positive ions will adsorb on the surface (red circles) and an electrical double layer (EDL) will develop within the liquid. The distribution of ions in the confined liquid is governed by the Nernst-Planck-Poisson-Boltzmann equations (see Methods, eqns. (M1)-(M3)). The steady-state concentration distribution of the ionic species depends on different modes of ion transport, namely (i) conduction along an electrostatic potential gradient, (ii) diffusion due to a concentration gradient, and (iii) advection induced by the evaporation-driven flows (convection). Any resulting imbalance in the ion distribution along the nanochannel leads to the measured electrical potential difference (Figure 1F). Overall, the total electrochemical potential of the i-th ionic component along the nanochannel length depends on the electrical potential, the ion concentration, and the chemical potential (see Methods eqn. (M4)). Interestingly, we observe that, due to the closed bottom surface of the nanochannel (z = 0), the studied geometry presents an intrinsic asymmetry in the surface-charge distribution. As a result, even in the absence of an evaporation-induced flow, an electrostatic potential, and therefore a $V_{OC}$ can be measured between the top and bottom surfaces of the nanochannel. Non-zero convection changes the distribution of ions and, depending on the flow profile, it can enhance or suppress the $V_{OC}$. Overall, our device structure presents three main advantages: (i) control over the liquid nanoconfinement, (ii) geometrical asymmetry within the nanochannel, and (iii) easy electrical contacting. Upon wetting, the $V_{OC}$ evolves with time starting from $V(t^0)$, which is zero (negative value) for moderate (high) concentrations and attains a stable steady state value $V(t^\infty)$ (Figure S3). The $V_{OC}$ as well as the current-voltage curves (I-V) of the device (Methods and Figure S3) were recorded as a function of the physical and chemical properties of the system and will be discussed extensively later.

**Modelling Approaches**

The development of a suitable model for these evaporation-driven HV devices is essential to interpret the experimental results and unravel the complex electrokinetic interactions. Based on the aforementioned observations, the studied system can be described with the equivalent electrical circuit shown in Figure 1E. It consists of three parts: 1) a current source, representing the evaporation-driven streaming current $I_{str}$, and the related ionic resistance, $R_{ionic}$, 2) a capacitor with a resistor in parallel, representing the EDL capacitance, $C_{EDL}$, and the associated charge transfer resistance, $R_{ct}$, due to geometrical asymmetrical and 3) an external load resistance, $R_L$. One can express the measured voltage at a given load resistance as follows:

$$V_L = \frac{\left[I_{str} + \frac{\sigma}{R_{ct}C_{edl}}\right] R_{ionic}}{1 + \frac{R_{ionic}}{R_L}} \qquad \mathbf{(1)}$$

For the open circuit condition ($R_L \rightarrow \infty$), we thus obtain:

$$V_{OC} = \left[I_{str} + \frac{\sigma}{R_{ct}C_{edl}}\right] R_{ionic} \qquad (\mathbf{2})$$

Concurrently, the power to an external load can be calculated as $P_L = V_L^2/R_L$ and the maximum power output $(\mathrm{dP_L})/(\mathrm{dR_L}) = 0$ is thus:

$$P_{max} = \frac{V_{OC}^2}{4R_{ionic}} \qquad (\mathbf{3})$$

The streaming current and the ionic resistance can be calculated from the ion concentration and the flow velocity (see Methods, eqns. (M5) and (M6)). However, no analytical expression is known for EDL capacitance for the geometry shown in Figure 1B. Therefore, we also developed a 3D numerical model (COMSOL®) to solve the Nernst-Planck-Poisson equation to determine the equilibrium distribution of ions and resulting electrostatic potentials (see Methods). To perform these calculations, however, we first need to identify an equivalent simplified geometry. Considering a top-view of the Si NPs array (Figure 1c), we observe that it is possible to define a hydraulic diameter ($D_h$) for the resulting nanochannels, with $D_{min}$ as the minimum separation between adjacent NPs as well as $A_{S-L}$, solid-liquid interface area normalized by sample area as:

$$D_h = \frac{2\sqrt{3}p^2}{\pi D_{np}} - D_{np} \qquad (\mathbf{4A})$$

$$A_{S-L} = 1 + \frac{\pi D_{np} L_{np}}{\sqrt{3}p^2} \qquad (\mathbf{4B})$$

$$D_{min} = p - D_{np} \qquad (\mathbf{4C})$$

The 3D array of nanochannels can thus be reduced to an equivalent annular nanochannel geometry of inner and outer diameters $R_1$ and $R_2$, respectively, such that $D_h$ and $A_{S-L}$ are equal to those of the original array (Figure 1C, D, and S1). A spatially uniform surface charge is used as a boundary condition. Importantly, as described by the Grahame Equation, its magnitude is governed by the diffuse layer potential and therefore by the chemical equilibrium between the silica surface and the electrolyte[35] (see Methods or Figure S4 ). This means that σ depends on the EDL thickness and therefore the bulk electrolyte concentration. The evaporative flux was instead used to define the mass flow rate in the channel. The computed potential difference between the top and bottom of the nanochannel is equal to the $V_{OC}$ and can be directly compared to experimental results (Figure 1D, F).

**Role of surface-charge and ion transport at low-to-moderate concentrations**

We first measure $V_{OC}$ as a function of ion concentration by using KCl solutions ranging from 1 µM to 0.1 M (Figure 2A). Indeed, this parameter plays a critical role both on the physical extension of the EDL and the chemical equilibrium at the interface. Regardless of the geometrical properties of the sample, our measurements clearly show a non-monotonic $V_{OC}$ trend as a function of electrolyte concentration. In particular, we observe a $V_{OC}$ peak for intermediate concentration values (0.1 – 1 mM), which can be more or less pronounced depending on the sample. This result is in sharp contrast to the monotonic decrease in $V_{OC}$ predicted by the mean-field Poisson-Boltzmann theory with a constant surface charge condition[25] (Figure S5). Instead, using our COMSOL model that accounts for the chemical equilibrium at the solid/liquid interface, it is possible to correctly capture the experimental trend (Figure 2B). In particular, for a fixed value of Γ in the chemical equilibrium equation, we retrieve the appearance of the $V_{OC}$ peak in the intermediate range of concentrations. Interestingly, the model reveals that the $V_{OC}$ peak magnitude increases for increasing the length of the Si NPs and for decreasing Si NPs size (i.e., increasing nanochannel size), consistent with experimental results. Additionally, increasing the value of Γ from 8 to 12 results in a more pronounced $V_{OC}$ peak. This is due to the increase in the number of available surface sites that can readily dissociate to enhance the surface charge density as schematically shown in Figure 2C. Overall, the observed trend for $V_{OC}$ is explained based on the interplay between the magnitude of the surface charge density and the extent of electrostatic screening (Figure 2C), quantified by the Debye length ($\lambda_D$). This is defined as $\lambda_\mathrm{D}^2 = \epsilon \mathrm{k_B T} / 4\pi \mathrm{e}^2 \mathrm{C_b}$ and is inversely proportional to the square root of the bulk ionic concentration ($C_b$). Thus, it monotonically decreases with increasing concentration. On the other hand, because of the chemical equilibrium at the interface, the magnitude of surface charge density and the diffuse layer potential both increase with concentration (Figure S4). Hence, a relatively high surface charge density and a moderate electrostatic screening give rise to an optimum condition and a peak in $V_{OC}$ at intermediate concentrations. To elucidate the significance of surface charge, we measured $V_{OC}$ for a single device at 1 µM concentration of KCl but varying the pH of the electrolyte across the isoelectric point by adding different concentrations of HCl (see time trace in Figure S6). As shown in Figure 2D, we observe the decrease in the magnitude of $V_{OC}$ with pH and then the change in sign below the isoelectric point. This highlights that the $V_{OC}$ is directly correlated to the surface charge, as the silanol groups at the surface could exist in one of these (*Si-O⁻, Si-OH, SiOH₂⁺*) forms depending on the pH, thus modulating the surface charge density and sign.

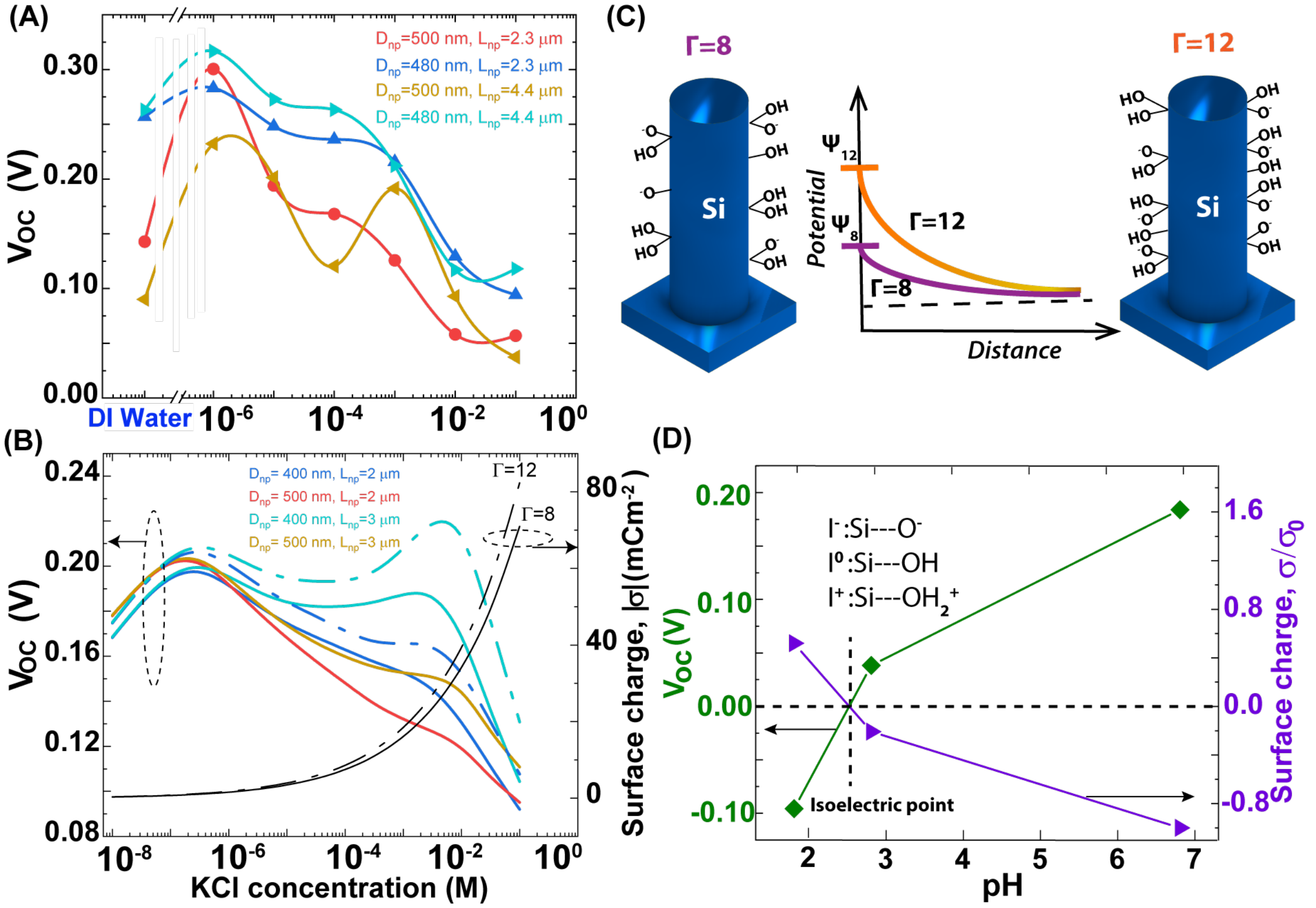

**Figure 2: Effect of Low to Moderate Electrolyte Concentrations. A)** Dependence of the measured $V_{OC}$ on the electrolyte concentration for a series of samples with different diameters ($D_{np}$) and lengths ($L_{np}$) of the NP. Lines are only a guide to the eye. **B)** Dependence of the simulated $V_{OC}$ on the electrolyte concentration. Two different values of the free parameter gamma (the number of available sites on silica surface) are used to show the importance of surface charge modulation (solid lines Γ = 8, dashed lines Γ = 12). Lines are only a guide to the eye. **C)** Schematic of the individual NP, showing a qualitative difference in surface charge density for **Γ = 8** *and* **Γ = 12.** The middle graph shows a qualitative difference in the surface potential $\psi$ and decay of EDL potential due to a change in surface charge density. **D)** Measured $V_{OC}$ at 1µM concentration for different pH values. The flipping of the sign of the surface charge across the isoelectric point is reflected in the sign of the measured $V_{OC}$. Depending on the pH, the magnitude of $I^-$, $I^+$, and $I^0$ changes that modulate the sign and magnitude of surface charge.

The values of $V_{oc}$ for deionized (DI) water are also shown in Figure 2A (left-most point in graph). DI water in standard condition has a hydronium ion concentration of $10^{-7}$ M and therefore the Debye length is large compared to $10^{-6}$ M KCl. Thus, from the perspective of EDL thickness, we might expect that DI water should give the highest $V_{OC}$. Interestingly, we observe that, in the case of DI water, $V_{OC}$ is much less than for 1μM electrolyte. This observation is consistent with previous reports although an accurate explanation has not been provided yet. From our equivalent circuit model, we note that the open circuit voltage is directly proportional to the ionic resistance (Eqn. (2)). $R_{ionic}$, in turn, is inversely proportional to the mobility of ions. Thus, higher (lower) mobility ions are expected to result in lower (higher) open circuit voltages. In particular, for the studied system with a negative surface charge, we anticipate a stronger dependence on the cation mobility. Thus, in the case of DI water, the hydronium ions dominate the ionic transport, which results in much lower ionic resistance. These have much higher mobility due to the hopping mechanism[37] of ion transport compared to the potassium cations used in the electrolyte, leading to a sharp decrease in $V_{OC}$.

To gain a deeper understanding of the effect of ionic mobility, we selected three chloride salts, namely LiCl, KCl, and CsCl, whose cations have increasing mobilities ($Li^+ < K^+ < Cs^+$) of 3.6, 7.32, 8.2 x$10^{-8}$ $m^2/Vs$ in aqueous medium[23]. For each salt, we varied the concentration from 1μM to 0.1M and recorded the $V_{OC}$ using the same device (Figure 3A). Both our experimental data and numerical modelling results (Figure 3B), which are in excellent agreement, clearly show that $V_{OC}$ increases with decreasing ion mobility, for the studied range of concentrations. Interestingly, we observe that the ionic mobility also influences the $V_{OC}$ peak at intermediate concentrations, which we previously related to the concentration dependent surface charge. We want to further highlight an interesting observation that connects the dependence of ionic mobility and the length of NP. The increase in length of NP and decrease in ionic mobility, both result in more pronounced intermediate peaks, which increase with surface charge, as shown in Figure 2A, B and Figure 3A, B. In particular, the lowest mobility cation ($Li^+$) exhibits the strongest $V_{OC}$ peak. Next, to quantify the difference in ion mobility as well as the contribution of cations and anions to $R_{ionic}$, we employed electrochemical impedance spectroscopy measurements for 1mM concentration of LiCl, KCl, and CsCl as well as KCl, KBr, KI. Figure 3C shows the diffusivity, and therefore the ionic mobility, determined by fitting the Nyquist plots using an appropriate electrical circuit (Figure S7). Interestingly, the large variation in diffusivity for different cations compared to the case of different anions confirms that cations dominate the ion transport in the studied system, as expected for a system with negative surface charge.

Next, we measure the I-V characteristic of the device (Figure S8) and determine the peak power ($P_{max}$) at each concentration (Figure 3D). We observe a non-monotonic power generation dependence on concentration, having a minimum of around 0.1 mM concentration. Based on the previous discussion, the increase in $P_{max}$ at higher (above 1 mM) concentrations is attributed to the increase in $R_{ionic}$, which dominates over the decrease in $V_{OC}$ as expressed in Eqn. (3). As a reference, we measured the bulk electrolyte conductivity for 0.01 mM, 1 mM, and 10 mM, as 1.64 µS, 95.4 µS, and 0.89 mS respectively. Importantly, we note that $R_{ionic}$ has contributions from both the bulk and surface conduction, which leads to a non-monotonic concentration dependent change in net ionic conductivity. This leads to an increase in power generation at low concentrations, which is attributed to the EDL overlap condition, where $R_{ionic}$ decreases due to an increase in surface conductivity. The relative contribution of the bulk and surface conductivity is governed by the concentration and size of the liquid confinement, which will be discussed in more detail in the last section.

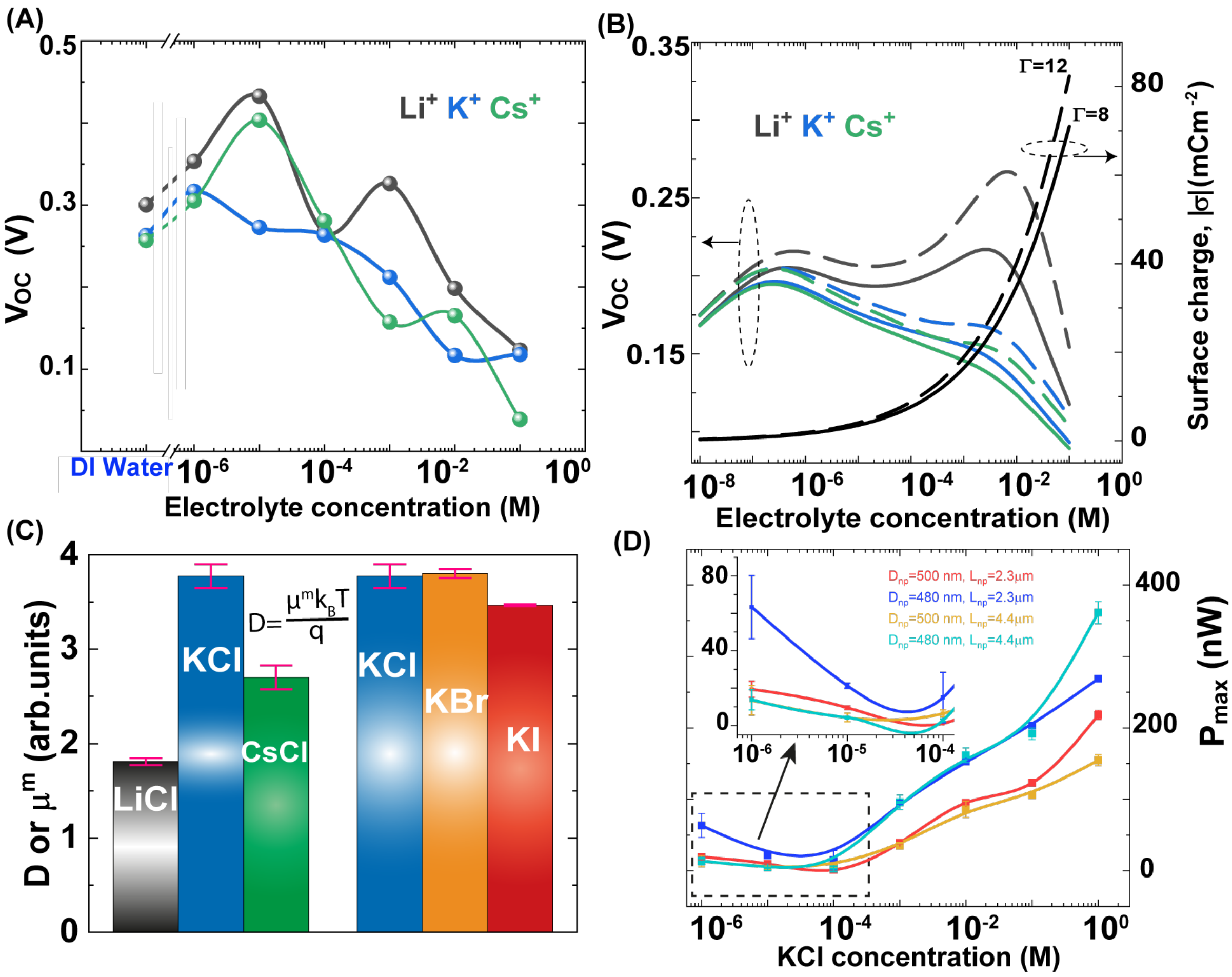

**Figure 3: Effect of ionic mobility on HV device performance at low-to-moderate concentration. A)** Measured concentration-varying $V_{OC}$ and its dependence on the counter-ion mobility. The studied electrolytes are LiCl (black curve), KCl (blue curve), and CsCl (green curve) in DI water. **B)** Calculated $V_{OC}$ as a function of electrolyte concentration for different ionic mobilities, following the order $Li^+ < K^+ < Cs^+$. Two different values of the free parameter Γ are used to show the variation with respect to the surface charge density. **C)** Diffusivity/mobility (related by Nernst-Einstein relation) of salts with different anions and cations determined from electrochemical impedance spectroscopy measurements for a constant concentration of 1mM. **D)** Measured maximum power output as a function of electrolyte concentration (KCl) for different diameters and lengths of the NPs. The colors correspond to the same devices as in Figures 2A and 4B.

## Effect of high concentration (>1M) on surface charge and ion transport

The Poisson-Boltzmann equation is based on mean field theory where one can ignore ion-ion correlations as well as the adsorption of ions to the surface[38,39]. It thus holds for low electrolyte concentrations (up to ~0.1 M), when $l_B^3 C_b \ll 1$, where $l_B = e^2/\varepsilon k_B T$ the Bjerrum length of the solution and $C_b$ is the bulk ion concentration, and ε is the dielectric constant of the electrolyte. However, for higher concentrations, the effect of ion-ion correlation becomes important. As a result, contrary to the Debye theory discussed above, the screening length ($k_L^{-1}$) increases with concentration and becomes dependent on the size of ions. In particular, for concentrated electrolytes, this dependence scales as[32] $k_L^{-1} \sim l_B C_b a^3$, where *a* is the diameter of the counter-ion. Furthermore, at concentrations above ~1M, the influence of ion adsorption becomes relevant, leading to a reduction in the effective surface charge. This effect therefore reverses the trend discussed earlier of increasing surface charge with increasing concentration. In particular, in the case of multivalent ions (Z ≥ 2), screening by counterions not only reduces the effective surface charge but can also reverse its sign.

This counterintuitive phenomenon is called *surface charge inversion* and arises due to the counterions forming a strongly correlated liquid (SCL)[34,40,41], where the electrostatic interactions are much larger than thermal fluctuations ($k_B T$). Surface charge inversion has been already observed for *Z* = 3 *and* 4[34]. Using pressure-driven streaming current measurements, it was also reported for Z = 2. However, there was an order of magnitude deviation from SCL theory in terms of the predicted minimum required ion concentration[33]. This deviation can be attributed to the omission of screening effects, which significantly enhances charge inversion[42]. Recently, direct measurements of surface excess charge using fluorescein particles have shown that at very high concentrations (>1M) even monovalent salts (Z=1) in silica confinement can give rise to charge inversion[43]. We thus explored experimentally the HV response of our devices for high electrolyte concentrations, including chloride salts of different-sized monovalent cations ($Li^+$, $K^+$, and $Cs^+$).

We show in Figure 4A the time traces of $V_{OC}$ for a few representative samples tested with 1M concentration KCl (purple and yellow curve), LiCl (green curve), and CsCl (pink curve). Upon wetting, we observe that *all* the curves exhibit an initial negative $V_{OC}^0$ value, where $V_{OC}^0 = V_{OC}(t=0)$, which was not observed for lower concentrations (see the gray curve for 1mM KCl concentration, and Figure S3). Interestingly, $V_{OC}^0$ becomes more negative with increasing salt concentration (from 0.5M up to 4M) as well as for decreasing the length of the Si NPs and increasing liquid nanoconfinement (Figure 4B. left). This can be directly related to the effect of counter-ion adsorption close to the Stern-plane leading to a reduction, or even inversion, of the surface charge. We note here that, based on Eqn. M8, a change in $V_{oc}$ sign must be related to surface charge inversion. Yet, a positive $V_{oc}$ cannot exclude such an effect due to the additional voltage term introduced by the asymmetrical device structures.

Considering a fixed surface charge of 1e/nm$^2$, which was measured previously for silicon HV structure[12], we can estimate the magnitude of the short-range interaction, which can give rise to charge inversion, as 2.7, 4.5, 8.8, and 14.3 $k_BT$ for ion valence Z=1, 2, 3, and 4 respectively[41]. This suggests that for monovalent cations, the high-concentration regions near the surface are mobile and are not expected to sustain charge inversion when equilibrium is achieved. Indeed, for some samples, the measured $V_{OC}$ changes over time, eventually stabilizing on a positive $V_{OC}^{\infty}$ value even at very high concentrations (Figure 4A, purple and pink curves, and Figure 4B (right)). In these cases, we can quantify the effective voltage generated by each sample as the difference between the initial and final $V_{OC}$ values, i.e. $V_{OC}^{eff} = V_{OC}^{\infty} - V_{OC}^{0}$. As a function of concentration, we observe a small but positive increase in $V_{OC}^{eff}$ (Figure S9), consistent with the increase in screening length at high concentration. Interestingly, we also repeatedly observed steady-state sign inversion of $V_{OC}$, which we postulate is related to charge inversion at high concentrations. (Figure 4A, green and yellow curves).

In order to overcome sample-to-sample variability and verify this observation (i.e., Figure 4A, yellow and purple curves), we first naively apply the SCL theory to calculate the critical concentration $C_0$[34]. For Z=1 and $\sigma \sim$1e/nm$^2$ we obtain $C_0 \sim 1M$. Based on this, we measured $V_{OC}^{\infty}$ for 4 different samples (S1-S4) at 1M electrolyte concentration using different cations like $Li^+$, $K^+$, and $Cs^+$ (Figure 4C). We observed that negative $V_{OC}^{\infty}$, and therefore charge inversion, is most likely in the case of small cations like $Li^+$, while no flipping in $V_{OC}^{\infty}$ sign was observed for big cations like $Cs^+$. This result is consistent with previous direct measurements of surface excess of fluorescein[43], which was positive for LiCl and NaCl (implying charge inversion of the silica surface) and the corresponding decay length at high concentrations. The dependence of ion size effect on charge reversal can be indeed attributed to EDL potential decay length at high concentrations, which scales as $\mathrm{k}^{-1} \sim \mathrm{l_B C_b a^3}$, and the specific adsorption that leads to excess cationic surface charge.

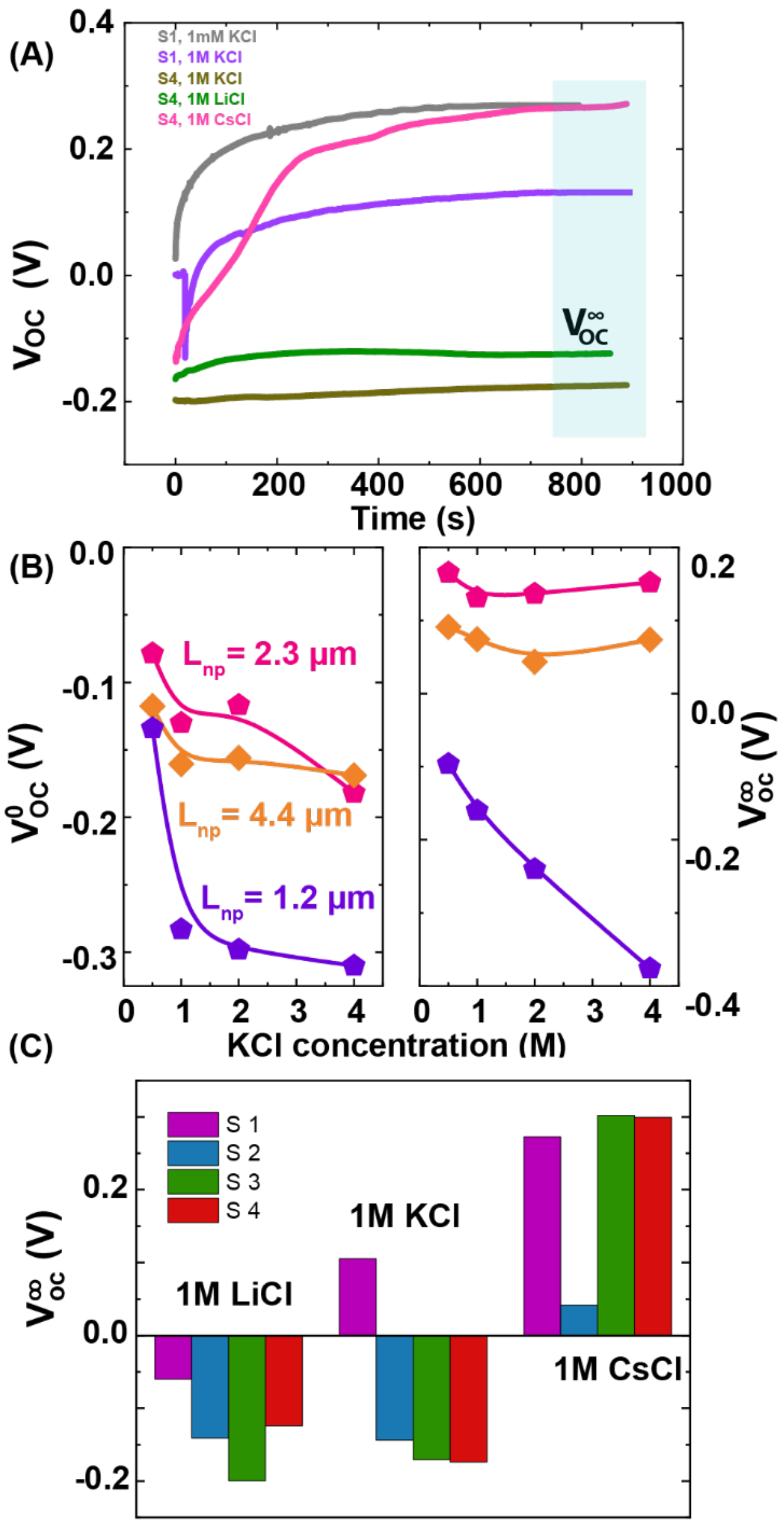

**Figure 4: Effect of electrolyte concentration on the measured $V_{OC}$ at high concentration. A)** Time-trace of $V_{OC}$ for various concentrations and geometrical parameters of the NP. In contrast to low concentration, the voltage at t=0 is negative instead of zero. **B)** Measured $V_{OC}$ at t=0 ($V_{OC}^{0}$, left) and steady-state $V_{OC}$, ($V_{OC}^{\infty}$, right) as a function of KCl concentration and for different samples (lines are a guide to the eye). **C)** Measured $V_{OC}^{\infty}$, for 4 different samples (S1-S4) of 1.2 µm length of the NPs and varying diameter of the NPs, at 1M concentration for different monovalent cations. Charge inversion (i.e. negative $V_{OC}^{\infty}$) is more likely to happen for small ions like ***$Li^+$*** than ***$Cs^+$***.

Overall, in terms of HV device performance for operation in saline conditions, like seawater or brine, these results suggest that specific electrokinetic interactions should be accounted for while engineering surface charge and device geometry. For example, based on the composition of the natural water[41], it would be important to engineer the interface properties to limit the surface charge density such that the critical concentration, $C_0$ is much higher than the concentration of ions in water. Furthermore, together with the result of the previous section, these measurements confirm that effective operation in high concentration solutions (i.e. brine) would also be possible.

**Effect of geometry and liquid nanoconfinement**

The impact of changes in surface charge and screening length on the device performance are strongly dependent on the size of liquid nanoconfinement, and therefore the geometry of the device. Although we have shown some selected results in the previous sections, we now analyze more in detail the effect of the mean nanopore diameter, $D_p$, and length, $L_{np}$. Firstly, we observe that the effect of geometrical parameters on measured $V_{OC}$ and $P_{max}$ is complex and non-linear (Figure 5). Indeed, changes in $D_p$ and $L_{np}$ modify the interfacial area ($A_{S\text{-}L}$). This, in turn, determines the net surface charge, the streaming current ($I_{str}$), and the associated ionic resistance ($R_{ionic}$, see Eqns. M4-5). In particular, based on Eqns. 4A-C, $A_{S\text{-}L}$ decreases with $D_p$ and increases with $L_{np}$, leading to an overall increase in $I_{str}$. Additionally, we observe that $R_{ionic}$ can be also expressed as: $R_{ionic} \equiv 1/\ (1/R_B + 1/R_S)$, where $R_B = L_{np}\ /(\pi D_p^2\ \sigma_B)$ is the bulk ionic resistance and $\sigma_B$ is the bulk conductivity, while $R_S$ is the surface resistance caused by the presence of the EDL and its distinct ionic conductivity. At small $D_p$ (or $D_{min}$), (Eqn. 4C), the EDL overlap increases, resulting in a higher space charge density and therefore a lower surface resistance contribution to $R_{ionic}$, as well as a higher streaming current.

Overall, the effect of surface resistance is quantified by a non-dimensional *Dukhin number*, $Du = \sigma_S/(\sigma_B D_p)$, and it has been shown to scale as[44] $Du \sim \sigma\lambda_D^2/eD_p$. This shows that the role of surface resistance becomes prominent in the case of small $D_p$ as well as for low ionic strengths. Hence, with the increase in electrolyte concentration, the bulk ionic resistance decreases, while the surface resistance can increase significantly, and therefore the relative contribution of the two is primarily determined by the diameter of the nanopore and bulk ionic concentration. From Figure 5A, B, we identify three different regimes with respect to change in the size of the nanoconfinement (i.e. $D_p$) in the following order: 1) EDL overlapping regime, where, for larger $D_p$ values, the space charge decreases leading to a lower $V_{OC}$ (blue shaded area). 2) Streaming current dominated regime, where an increase in $D_P$ leads to higher streaming current (red shaded area). 3) Ionic resistance dominated regime, where a decrease in ionic resistance limits $V_{OC}$ (grey shaded area) which can level off or even decrease. The concurrent variation in $P_{max}$ is due to the interplay between the ionic resistance and $V_{OC}$ as shown in Eqn. (4). It can be noted from our measurements that not all samples/electrolyte concentrations show the existence of all regimes. This is due to the limited number of samples with different geometrical parameters used for measurements and some intrinsic variability in the initial surface charge.

Finally, with an increase in the length of the NP, the ionic resistance increases, and therefore $V_{OC}$ is expected to increase too. However, for a very long etching time, the surface composition can change, towards a reduction of the available surface sites on the NP due to HF removal of the native oxide layer on Silicon. As a result, we observe a $V_{OC}$ saturation for large etching times, i.e., for longer Si NPs, as shown in Figure 5C. We would furthermore draw the readers' attention that other factors such as a change in the fluidic flow pattern and sticking of long NPs due to surface forces can also affect the electrokinetic response. However, the discussion related to those aspects is beyond the scope of this manuscript. Interestingly, if we consider the peak power per-unit solid-liquid interface area (Figure 5D) we observe that it decreases with an increase in interface area. This means that an optimum value of peak power exists for a given pitch and diameter of the NPs. Overall, this shows that geometrical parameter ($D_p$, $A_{S\text{-}L}$, and $L_{NP}$) plays a critical role in controlling the performance metric of the HV device which has a complex interplay with the surface charge. In particular, for optimizing the device performance one has to carefully identify the regime of operation for a given set of geometrical parameters and variability of surface charge depending on the fabrication process.

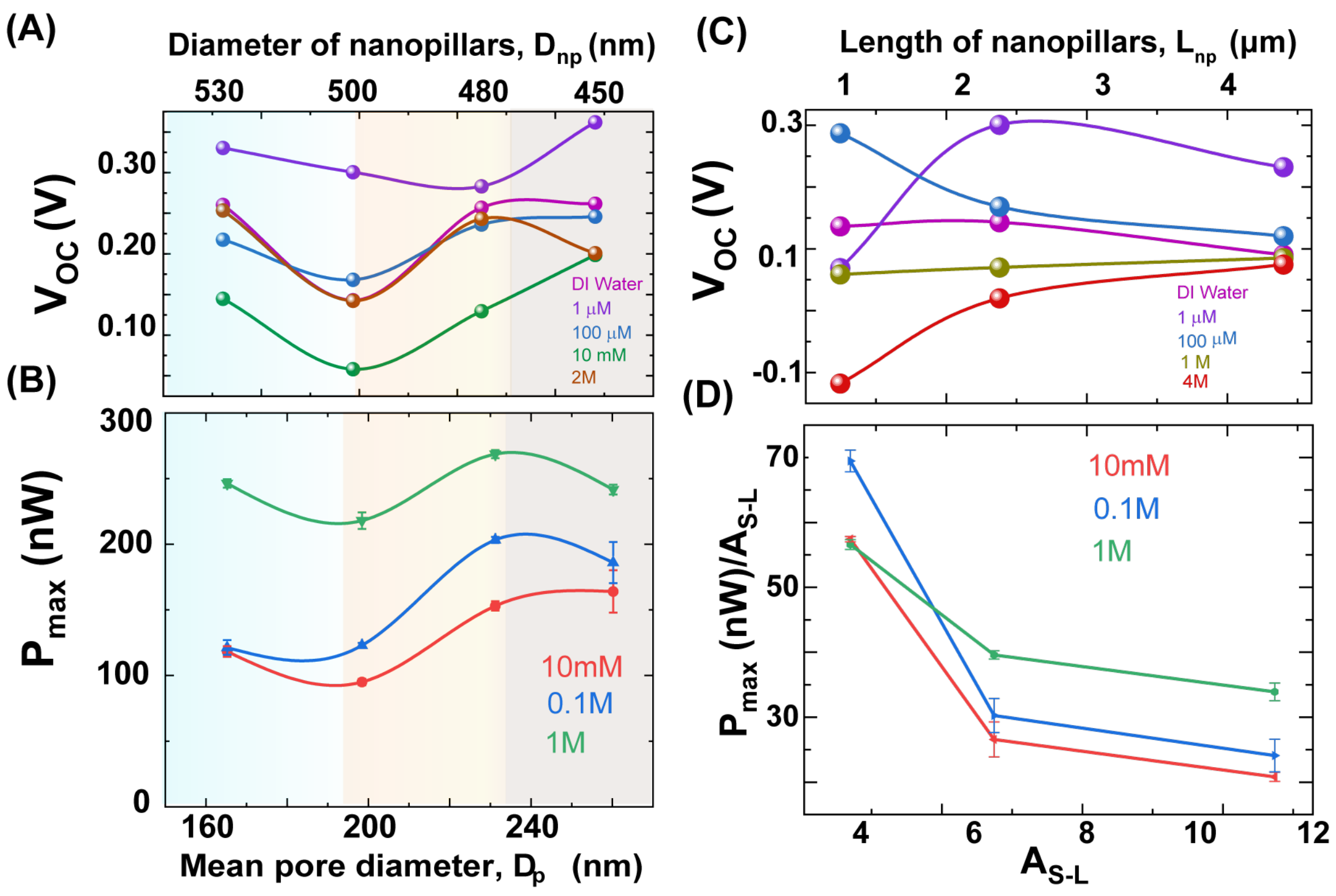

**Figure 5: Dependence of geometrical parameters on the measured electrical outputs. A)** Measured $V_{OC}$ as a function of the mean pore diameter ($D_p$) obtained by varying the diameters of the Si NP. Data are for a fixed length of the NP equal to 2.3 µm. **B)** Measured peak power as a function of the mean pore diameter for the same length of NP and various concentrations. The different colored regions represent the different regimes of ion transport as described in the text. **C)** Measured $V_{OC}$ as a function of the normalized solid-liquid interface area, $A_{S-L}$. This is varied by changing the length of the NP ($L_{np}$), as shown on the top axis, for a fixed NP diameter equal to 480 nm. **D)** Measured power normalized by solid-liquid interface area. The respective lengths of NPs are shown on the top axis.

## Conclusion

To summarize, we found that ion-concentration-dependent surface charge, together with ion mobility, entails optimal HV operating conditions that deviate from conventional low-concentration expectations. Additionally, we observed that HV devices can successfully operate at concentrations exceeding the critical value ($C_0$), charge inversion affecting the sign and magnitude of the steady-state $V_{oc}$. This leads us to conclude that to-date studies of evaporation-driven HV devices have been unnecessarily limited to deionized water or low ion concentrations (regime of EDL overlap for ~10-100nm liquid confinement size). Instead, our results show that there is ample margin for improving their performance and extending their scope within the wide range of salinity conditions available in natural and processed water. More broadly, our controlled experiments and unique quantitative modeling clarify the complex interplay of surface charge, liquid properties, and geometrical constraints in evaporation-driven HV systems. They indeed uncover how different combinations of solid and liquid physical properties define distinct limiting regimes of operation. In this regard, Figure 6A highlights four main controlling quantities we have identified, namely ion concentration (C), surface charge (σ), solid-liquid interface ($A_{solid-liquid}$), and liquid nanoconfinement ($D_p$), the latter two depending in our devices on the length ($L_{np}$) and diameter ($D_{np}$) of the Si NPs. It also shows how changes in the magnitude of these parameters, impact $V_{oc}$ and $P_{max}$. Importantly, within this complex and multidimensional parameter space, it is possible to take advantage of different interfacial phenomena and identify suitable operating conditions for different salinity values. We found that for fresh water conditions (~1 mM), EDL overlaps is necessary. This can be realized, as expected, under low total surface charge and small nanochannel size (high confinement). Yet, by increasing the surface area or the surface charge, larger nanochannel sizes become viable. Furthermore, under seawater conditions (from ~10mM up to ~100mM), thanks to the chemical equilibrium at the interface, by controlling the surface charge (i.e. Gamma value in Figure 2B) an optimum can be engineered at large $D_p$ values (> 100nm). This implies that nm-scale geometrical confinement can be avoided, simplifying the scalability of HV devices. Finally, at high salinity levels (above 1M) charge inversion can be leveraged by minimizing the solid-liquid interface and initial surface charge. Yet, long-term operation at very high concentrations can be challenging due to ion adsorption and salt crystallization, which directly affect the surface properties and geometry of the nanostructure (Figure S10). Thus, it will require further investigation.

From the perspective of device geometry engineering, our results confirmed that structural asymmetries lead to an open circuit voltage contribution entirely due electrostatic effect. Importantly, this term is what generates a sizable $V_{oc}$ (>0.1 V) in our system, despite very low evaporation rates (Figure S2). This could be further enhanced by engineering a spatially non-homogeneous surface charge distribution through chemical or physical processing[45,46]. Interestingly, taking advantage of our comprehensive model for predicting the performance of HV devices, we expect that the $V_{OC}$ performance metric can be further augmented by improving the rate of evaporation. Figure 6B shows that, depending on surface charge and geometry of the nanoconfinement, $V_{oc}$ can be doubled by a five-fold increase of the evaporation rate compared to ambient conditions, which means that power can be increased up to four times. This is largely due to the enhancement in streaming current and to be confirmed it will require a more in-depth understanding of the fluid dynamics in HV devices. Overall, based on these guidelines, by knowing the detailed composition of ions in the used water, it becomes possible to optimize geometrical and interfacial properties of the evaporation-driven HV devices. Thus, our work, which lies between nanofluidic studies of individual nanochannels and macro-scale porous device testing, offers critical insight into how to enhance the performance of evaporation-driven HV devices and points towards broader application opportunities for these self-powered systems.

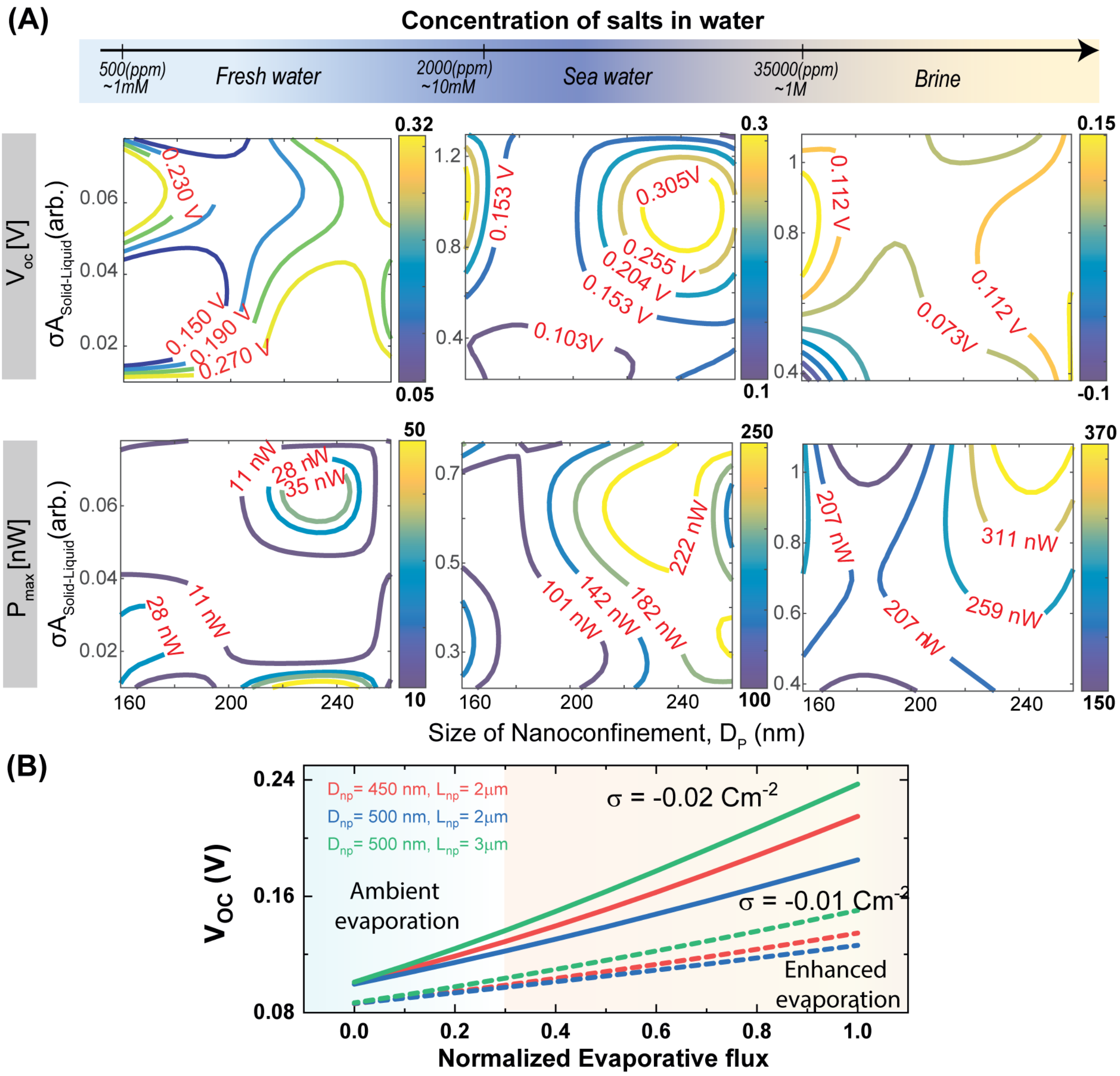

**Figure 6: Strategies towards boosting HV device performance. A)** Contour plots of the measured $V_{OC}$ and $P_{max}$ across a wide range of salinity conditions available in natural and processed water. At low salinity, optimal performance can be achieved by leveraging the EDL overlap. At intermediate salinity, optimal operation can be achieved by engineering the surface charge and geometrical parameters, which is also highlighted by the intermediate peaks in Figure 2A-B. At very high salinity, one can leverage the charge inversion regime for optimal operations of the device. **B)** Simulated open circuit voltage for a series of different geometrical parameters as a function of evaporative flux for two values of surface charge. This shows that the performance can be further augmented by fluid dynamic consideration and enhancing the rate of evaporation from the system.

## Methods

### Fabrication of Silicon nanopillars array

Metal-assisted chemical etching (MACE) of crystalline silicon combined with colloidal lithography was used for the fabrication of a cm-scale array of silicon NPs[47,48] (Figure S11). It involves the self-assembly of polystyrene nanospheres at the water-air interface. Then the non-closed-packed assembly of PS nanospheres was compressed to a pressure of approximately 25-30($N/m^2$) using the Langmuir-Blodgett system, which resulted in a homogeneous closed-packed hexagonal lattice of PS nanospheres[49]. The closed-packed monolayer was then transferred to Pirhanna-cleaned Silicon substrates which were diced into 2cm X 2cm chips. Plasma etching was used to reduce the diameter of the PS nanospheres with an initial diameter of $d$ = 600*nm*. After gold-sputtering deposition of thickness 20*nm* and lift-off, a gold nanomesh is formed which is used as an etching mask for MACE. Prior to gold sputtering, 3-5 nm of Ti is sputtered as an adhesion layer. This forms a stable contact between gold and the substrate to avoid delamination during MACE. The liftoff was done by putting the substrate in Toluene and ultrasonicating it at moderate power for 3-5 minutes at room temperature. Finally, MACE was performed by putting the substrate in an aqueous $HF/H_2O_2$ solution with a volumetric percentage of HF and $H_2O_2$ as 10% and 2% respectively. The diameter of the NPs is controlled by changing the time of plasma trimming of the polystyrene nanosphere, while the length of the NPs is controlled by changing the MACE time. The obtained sample was then treated with oxygen plasma (60 sec, 1000 W), to improve the hydrophilicity and surface charge.

### Electrical Measurements

The electrical measurements configuration is shown in Figure 1A, in which silicon NP is the active substrate and the Al layer acts as a back electrode while Ag/AgCl is used as a top electrode. For comparison, we tested the planar silicon device and NP device with Ag/AgCl electrodes and different porous top electrodes (Figure S12). The planar silicon gives almost zero $V_{OC}$. We used an Ag/AgCl electrode as it is considered fully reversible, which ensures that the charges accumulated in the electrode EDL are entirely consumed by the electrodes at overpotentials, which are practically zero, and there is no unwanted potential difference, which induces a conduction current. The open circuit I-V and EIS measurements were done using CHI bipotentiostat. During EIS measurement, for covering the entire area of the active Silicon NPs, a porous graphite electrode was used instead of Ag/AgCl. I-V characteristics were obtained using linear sweep voltammetry (-0.5-0.5V) at a scan rate of 0.01V/s. The EIS measurements were done at zero applied DC voltage with an amplitude of 10mV in the frequency range of 100Hz-1MHz.

**Modelling and simulation**

$$\nabla \cdot J_i + U \cdot \nabla c_i = 0 \qquad (\boldsymbol{M1})$$

$$J_i = -D\nabla c_i - z_i \mu_i^m F c_i \nabla \Psi \qquad (\boldsymbol{M2})$$

$$\nabla^2 \Psi = -\frac{1}{\epsilon_0 \epsilon_r} \sum_i F z_i c_i exp\left(-\frac{e z_i \Psi}{k_B T}\right) \qquad (\boldsymbol{M3})$$

For modeling the evaporation-driven electrokinetic conversion phenomenon in an array of vertical NP, we considered different modes of ion transport using the Nernst-Planck equation for the transport of dilute species, coupled with the Poisson-Boltzmann equation for equilibrium distribution of ions. The details of the simulation such as package use, and boundary conditions are given in S13. Diffusion becomes relevant because only the bottom part of the nanopores has surface charge, streaming of ions due to evaporation-driven flow, and migration of ions due to EDL potential. Based on the potential and charge distribution, it is also possible to compute the streaming current and ionic resistance to be used in the equivalent electrical circuit model:

$$I_{str} = \int_A \sum_i e z_i n_i(r) U(r) dA \qquad (\boldsymbol{M4})$$

$$\frac{1}{R_{ionic}} = \int_A \frac{1}{L} \sum_i e \mu_i^m n_i(r) + \frac{\epsilon_0 \epsilon_r}{\eta L} (\Psi(r) - \Psi_0) dA \qquad (\boldsymbol{M5})$$

Performing a full 3D simulation for a hexagonal lattice is computationally expensive, so we simplified our simulation, by transforming it to an equivalent annular cylindrical geometry. During this transformation, the two important parameters, the hydraulic diameter, and the solid-liquid interfacial area were kept constrained which gives us unique inner and outer radii $R_1$ and $R_2$ respectively. An evaporative flux is used to impose the flow rate condition (Figure S2). A spatially uniform surface charge was used as a boundary condition.

$$\Psi_d(\sigma) = \frac{k_B T}{e}\left[ln\left(\frac{-\sigma}{\sigma + e\Gamma}\right) - \ln(10)(pH - pK)\right] - \frac{\sigma}{C} \qquad (\boldsymbol{M6})$$

$$\sigma\,(\psi_d) = \frac{2\epsilon_0 \epsilon \kappa k_B T}{e}\left[\sinh\frac{e\psi_d}{2k_B T} + \frac{2}{\kappa r}\tanh\frac{e\psi_d}{4k_B T}\right] \qquad (\boldsymbol{M7})$$

The magnitude of surface charge was governed by the equilibrium between the silica surface and the electrolyte M6. In the case of an isolated surface with a curvature or non-overlapping double layer, the charge density on the surface satisfies the Grahame equation[35] M7, and by solving this equation coupled with the M6, we can obtain surface charge as a function of electrolyte concentration. In our model, we use Gamma as a free parameter as it can vary largely depending on the surface preparation. So here, we use different values of Gamma based on the available literature to show how the number of surface sites available for deprotonation affects surface change, and therefore the $V_{OC}$. We did not attempt to retrieve the experimental curve using our simulation, because we used the Grahame equation together with chemical equilibrium boundary conditions that hold for isolated surfaces. No analytical solution is available for a non-isolated surface and in that case, one has to sweep sigma for a range of values of surface potential and then find the intersection point with the chemical equilibrium condition.

To give a theoretical description for geometrical asymmetry-induced potential difference we define the total electrochemical potential of the i-th ionic component, $\Phi_i$, along the nanochannel length ($z$-direction), which is given by:

$$\Phi_i(z) = \Phi_i^0 + k_B T \ln\left(\frac{c_i(z)}{c_i^0}\right) + e z_i \Psi_i(z) + \mu_{i,ex} \qquad (\boldsymbol{M8})$$

where $\Psi_i$ is the electrical potential, determined by the overall charge distribution, where $\Phi^0_i$, $c^0_i$ , $z_i$ , $\mu_{i,\,ex}$ are the standard chemical potential, standard concentration, valence, and excess chemical potential[50] of the i-th component, respectively. At equilibrium, the total electrochemical potential of the system must be the same, at all spatial locations, so $\sum_i \Phi_i(z=0) = \sum_i \Phi_i(z=L)$ . The resulting potential difference between the top and bottom of the nanoconfinement is equal to the $V_{OC}$ due to the difference in the distribution of ionic concentration (Figure S14).

**Acknowledgment**

The authors acknowledge the support of the Swiss National Science Foundation (SNSF) through the Korean-Swiss Science and Technology Cooperation Fund (Grant No. IZKSZ2_188341), and the Swiss Government Excellence fellowship. The authors also acknowledge the support of the following experimental facilities at EPFL: Center of MicroNanoTechnology (CMi) and Interdisciplinary Centre for Electron Microscopy (CIME).

**Supporting Information Available**

All supplemental details related to fabrication, experimental data, calculation, and simulation can be found in the Supporting Information file. In addition a supplementary video is available showing the HV testing.

**References**

1. Zhang, Z. *et al.* Emerging hydrovoltaic technology. *Nature Nanotech* **13**, 1109–1119 (2018).
2. Yin, J., Zhou, J., Fang, S. & Guo, W. Hydrovoltaic Energy on the Way. *Joule* **4**, 1852–1855 (2020).
3. Cavusoglu, A.-H., Chen, X., Gentine, P. & Sahin, O. Potential for natural evaporation as a reliable renewable energy resource. *Nature Communications* **8**, 617 (2017).
4. Cavusoglu, A.-H. A Theory of Renewable Energy from Natural Evaporation. (Columbia University, 2017). doi:10.7916/D8DJ5SXS.
5. Gao, F., Li, W., Wang, X., Fang, X. & Ma, M. A self-sustaining pyroelectric nanogenerator driven by water vapor. *Nano Energy* **22**, 19–26 (2016).
6. Zhu, L., Gao, M., Peh, C. K. N., Wang, X. & Ho, G. W. Self-Contained Monolithic Carbon Sponges for Solar-Driven Interfacial Water Evaporation Distillation and Electricity Generation. *Advanced Energy Materials* **8**, 1702149 (2018).

7. Zhang, X. *et al.* Conversion of solar power to chemical energy based on carbon nanoparticle modified photo-thermoelectric generator and electrochemical water splitting system. *Nano Energy* **48**, 481–488 (2018).

8. Zhu, L., Ding, T., Gao, M., Peh, C. K. N. & Ho, G. W. Shape Conformal and Thermal Insulative Organic Solar Absorber Sponge for Photothermal Water Evaporation and Thermoelectric Power Generation. *Advanced Energy Materials* **9**, 1900250 (2019).

9. Das, S. S., Pedireddi, V. M., Bandopadhyay, A., Saha, P. & Chakraborty, S. Electrical Power Generation from Wet Textile Mediated by Spontaneous Nanoscale Evaporation. *Nano Lett.* **19**, 7191–7200 (2019).

10. Shuvra Das, S., Kar, S., Anwar, T., Saha, P. & Chakraborty, S. Hydroelectric power plant on a paper strip. *Lab on a Chip* **18**, 1560–1568 (2018).

11. Xue, G. *et al.* Water-evaporation-induced electricity with nanostructured carbon materials. *Nature Nanotech* **12**, 317–321 (2017).

12. Qin, Y. *et al.* Constant Electricity Generation in Nanostructured Silicon by Evaporation-Driven Water Flow. *Angewandte Chemie* **132**, 10706–10712 (2020).

13. Lü, J., Ren, G., Hu, Q., Rensing, C. & Zhou, S. Microbial biofilm-based hydrovoltaic technology. *Trends in Biotechnology* **41**, 1155–1167 (2023).

14. Sun, J. *et al.* Electricity generation from a Ni-Al layered double hydroxide-based flexible generator driven by natural water evaporation. *Nano Energy* **57**, 269–278 (2019).

15. Li, J. *et al.* Surface functional modification boosts the output of an evaporation-driven water flow nanogenerator. *Nano Energy* **58**, 797–802 (2019).

16. Shao, B. *et al.* Electron-Selective Passivation Contacts for High-Efficiency Nanostructured Silicon Hydrovoltaic Devices. *Advanced Materials Interfaces* **8**, 2101213 (2021).

17. Shao, B. *et al.* Bioinspired Hierarchical Nanofabric Electrode for Silicon Hydrovoltaic Device with Record Power Output. *ACS Nano* **15**, 7472–7481 (2021).

18. Shao, B. *et al.* Boosting electrical output of nanostructured silicon hydrovoltaic device via cobalt oxide enabled electrode surface contact. *Nano Energy* **106**, 108081 (2023).

19. Liu, C. *et al.* Hydrovoltaic energy harvesting from moisture flow using an ionic polymer–hydrogel–carbon composite. *Energy Environ. Sci.* **15**, 2489–2498 (2022).

20. Xin, X. *et al.* Hydrovoltaic effect-enhanced photocatalysis by polyacrylic acid/cobaltous oxide–nitrogen doped carbon system for efficient photocatalytic water splitting. *Nat Commun* **14**, 1759 (2023).

21. Dao, V.-D., Vu, N. H., Thi Dang, H.-L. & Yun, S. Recent advances and challenges for water evaporation-induced electricity toward applications. *Nano Energy* **85**, 105979 (2021).

22. Jiang, Z. *et al.* Simultaneous electricity generation and steam production from a wide range of salinity by using unique nanofluidic diode. *Nano Energy* **108**, 108220 (2023).

23. van der Heyden, F. H. J., Bonthuis, D. J., Stein, D., Meyer, C. & Dekker, C. Electrokinetic Energy Conversion Efficiency in Nanofluidic Channels. *Nano Lett.* **6**, 2232–2237 (2006).

24. Ren, Y. & Stein, D. Slip-enhanced electrokinetic energy conversion in nanofluidic channels. *Nanotechnology* **19**, 195707 (2008).

25. van der Heyden, F. H. J., Stein, D. & Dekker, C. Streaming Currents in a Single Nanofluidic Channel. *Phys. Rev. Lett.* **95**, 116104 (2005).

26. van der Heyden, F. H. J., Bonthuis, D. J., Stein, D., Meyer, C. & Dekker, C. Power Generation by Pressure-Driven Transport of Ions in Nanofluidic Channels. *Nano Lett.* **7**, 1022–1025 (2007).

27. Stein, D., Kruithof, M. & Dekker, C. Surface-Charge-Governed Ion Transport in Nanofluidic Channels. *Phys. Rev. Lett.* **93**, 035901 (2004).

28. Feng, J. *et al.* Single-layer MoS2 nanopores as nanopower generators. *Nature* **536**, 197–200 (2016).

29. Light-Enhanced Blue Energy Generation Using MoS2 Nanopores - ScienceDirect. https://www.sciencedirect.com/science/article/pii/S2542435119302090.

30. Xiao, K., Jiang, L. & Antonietti, M. Ion Transport in Nanofluidic Devices for Energy Harvesting. *Joule* **3**, 2364–2380 (2019).

31. Principles and applications of nanofluidic transport | Nature Nanotechnology. https://www.nature.com/articles/nnano.2009.332.

32. Lee, A. A., Perez-Martinez, C. S., Smith, A. M. & Perkin, S. Scaling Analysis of the Screening Length in Concentrated Electrolytes. *Phys. Rev. Lett.* **119**, 026002 (2017).

33. van der Heyden, F. H. J., Stein, D., Besteman, K., Lemay, S. G. & Dekker, C. Charge Inversion at High Ionic Strength Studied by Streaming Currents. *Phys. Rev. Lett.* **96**, 224502 (2006).

34. Besteman, K., Zevenbergen, M. A. G. & Lemay, S. G. Charge inversion by multivalent ions: Dependence on dielectric constant and surface-charge density. *Phys. Rev. E* **72**, 061501 (2005).

35. Behrens, S. H. & Grier, D. G. The charge of glass and silica surfaces. *The Journal of Chemical Physics* **115**, 6716–6721 (2001).

36. Morita, M., Ohmi, T., Hasegawa, E., Kawakami, M. & Ohwada, M. Growth of native oxide on a silicon surface. *Journal of Applied Physics* **68**, 1272–1281 (1990).

37. Agmon, N. The Grotthuss mechanism. *Chemical Physics Letters* **244**, 456–462 (1995).

38. Borukhov, I., Andelman, D. & Orland, H. Steric Effects in Electrolytes: A Modified Poisson-Boltzmann Equation. *Phys. Rev. Lett.* **79**, 435–438 (1997).

39. Borukhov, I., Andelman, D. & Orland, H. Adsorption of large ions from an electrolyte solution: a modified Poisson–Boltzmann equation. *Electrochimica Acta* **46**, 221–229 (2000).

40. Besteman, K., Zevenbergen, M. A. G., Heering, H. A. & Lemay, S. G. Direct Observation of Charge Inversion by Multivalent Ions as a Universal Electrostatic Phenomenon. *Phys. Rev. Lett.* **93**, 170802 (2004).

41. Shklovskii, B. I. Screening of a macroion by multivalent ions: Correlation-induced inversion of charge. *Phys. Rev. E* **60**, 5802–5811 (1999).

42. Nguyen, T. T., Grosberg, A. Yu. & Shklovskii, B. I. Macroions in Salty Water with Multivalent Ions: Giant Inversion of Charge. *Phys. Rev. Lett.* **85**, 1568–1571 (2000).

43. Gaddam, P. & Ducker, W. Electrostatic Screening Length in Concentrated Salt Solutions. *Langmuir* **35**, 5719–5727 (2019).

44. Sarkadi, Z., Fertig, D., Valiskó, M. & Boda, D. The Dukhin number as a scaling parameter for selectivity in the infinitely long nanopore limit: Extension to multivalent electrolytes. *Journal of Molecular Liquids* **357**, 119072 (2022).

45. Nanofluidic Diode. https://pubs.acs.org/doi/epdf/10.1021/nl062924b doi:10.1021/nl062924b.

46. Constantin, D. & Siwy, Z. S. Poisson-Nernst-Planck model of ion current rectification through a nanofluidic diode. *Phys. Rev. E* **76**, 041202 (2007).

47. Wendisch, F. J., Rey, M., Vogel, N. & Bourret, G. R. Large-Scale Synthesis of Highly Uniform Silicon Nanowire Arrays Using Metal-Assisted Chemical Etching. *Chem. Mater.* **32**, 9425–9434 (2020).

48. Kheyraddini Mousavi, B. *et al.* Metal-assisted chemical etching of silicon and achieving pore sizes as small as 30 nm by altering gold thickness. *Journal of Vacuum Science & Technology A* **37**, 061402 (2019).

49. Thangamuthu, M., Santschi, C. & Martin, O. J. F. Reliable Langmuir Blodgett colloidal masks for large area nanostructure realization. *Thin Solid Films* **709**, 138195 (2020).

50. Bazant, M. Z., Kilic, M. S., Storey, B. D. & Ajdari, A. Towards an understanding of induced-charge electrokinetics at large applied voltages in concentrated solutions. *Advances in Colloid and Interface Science* **152**, 48–88 (2009).

# Supplementary for: Salinity-Dependent Interfacial Phenomena Towards Hydrovoltaic Device Optimization

Tarique Anwar and Giulia Tagliabue*

*Laboratory of Nanoscience for Energy Technologies (LNET), STI, École Polytechnique Fédérale de Lausanne (EPFL), Lausanne 1015, Switzerland*

*Email: giulia.tagliabue@epfl.ch

## S1: SEM Image statistics

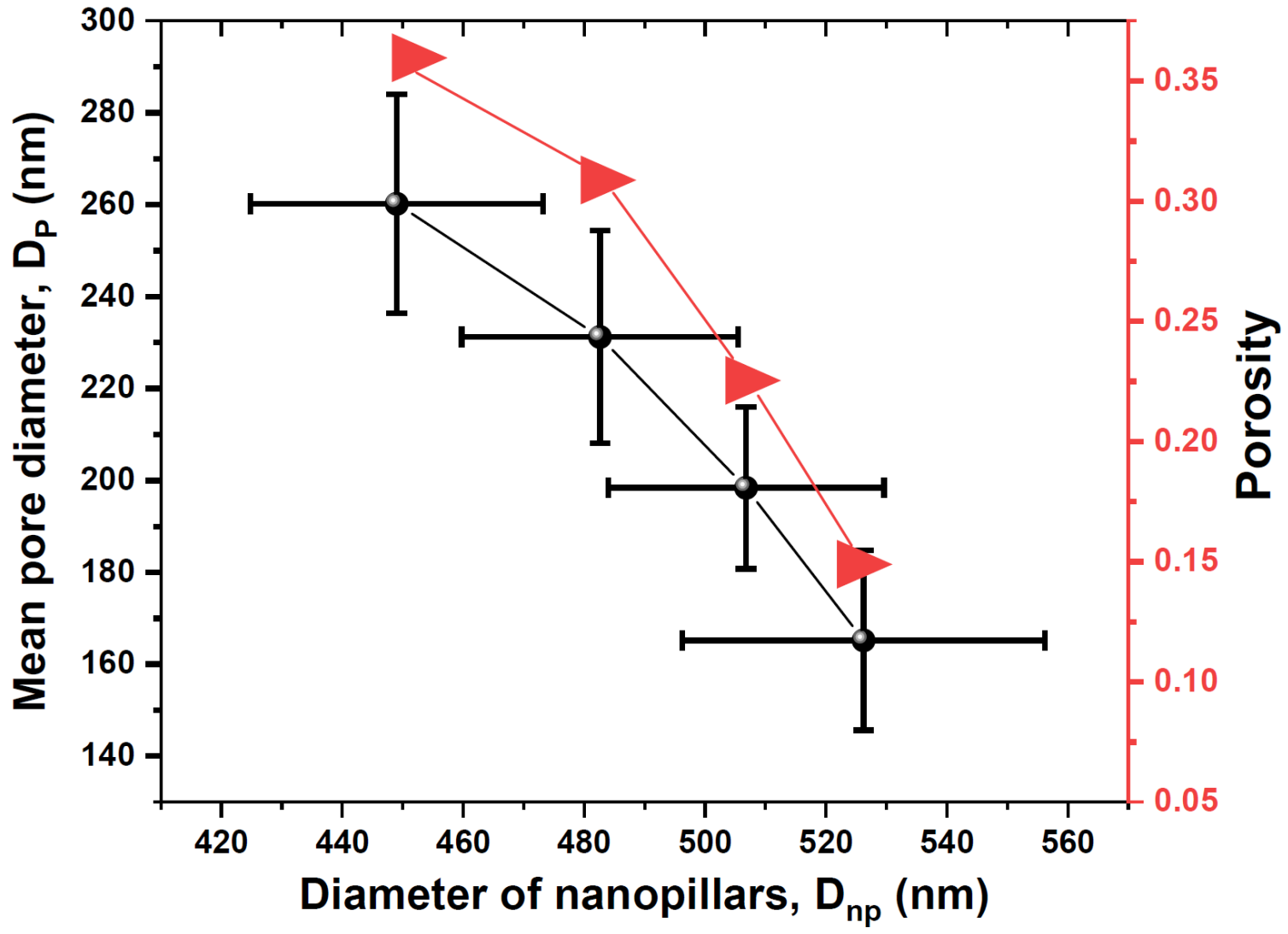

**Figure S1:** The SEM image of different samples used in the experiments was used for analysis to obtain the mean pore diameter and porosity as a function of Nanopillar diameter.

Si NP of different diameters was obtained by changing the plasma etching time of the polystyrene nanosphere from 0-15 minutes. The final diameter of the NP was non-linear with respect to the plasma etching time. This was probably due to the etching of the Nano mesh while performing MACE resulting in reduced diameter. It can also be seen that the diameter of the NP without any plasma etching of the polystyrene nanosphere is less than 600 nm (the initial diameter). On the other hand, different Si NPs of different length was obtained by varying the MACE time from 5 to 20 minutes. The length of pillars was almost linearly increasing with MACE time. For MACE time of 5, 10, and 20 minutes, the lengths of the pillars were *1.23μm, 2.3μm, and 4.4μm* respectively. The longer etching time not only increases the length of the NP but also affects the surface composition such as roughness, change in surface activity, or surface charge regulation effects. This could have a significant effect on the $V_{OC}$, as can be seen from simulation **Fig. 2B** that a small change in the free parameter **Γ** can affect the surface charge and give a much different $V_{OC}$.

**S2: Measured evaporative flux and the corresponding mass flow rate imposed in the simulation**

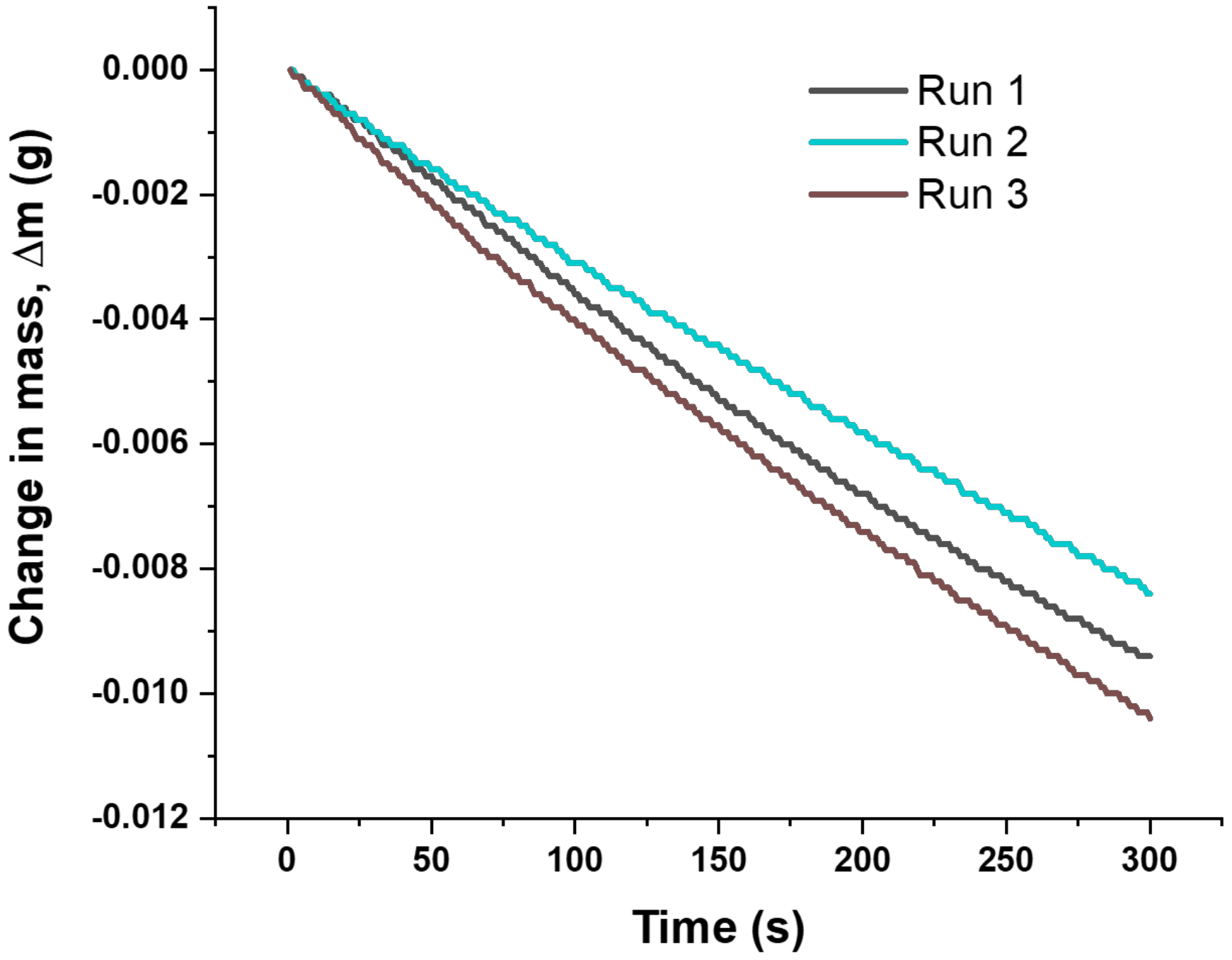

**Figure S2:** The change in mass due to evaporation in an ambient environment was recorded using a microbalance. Three successive measurements were done on the same sample after thorough drying. The average value of ambient evaporative flux was estimated to be around $2 \times 10^{-5} gcm^{-2}s^{-1}$ $or$ $\sim 1$ $\boldsymbol{kgm^{-2}h^{-1}}$**.** The flow rate imposed in the simulation was normalized with respect to the porosity of the sample in the range of 0.1-0.3 as shown in **Fig. S1**. In the simulation, the parabolic flow profile (for pressure driven) and constant flow profile (limiting case of electroosmotic flow for low Debye-length) were considered, and the resultant $V_{OC}$ was almost invariant with respect to the chosen profile under the range of evaporative flux used.

## S3: Voltage time for different geometrical parameters and concentration

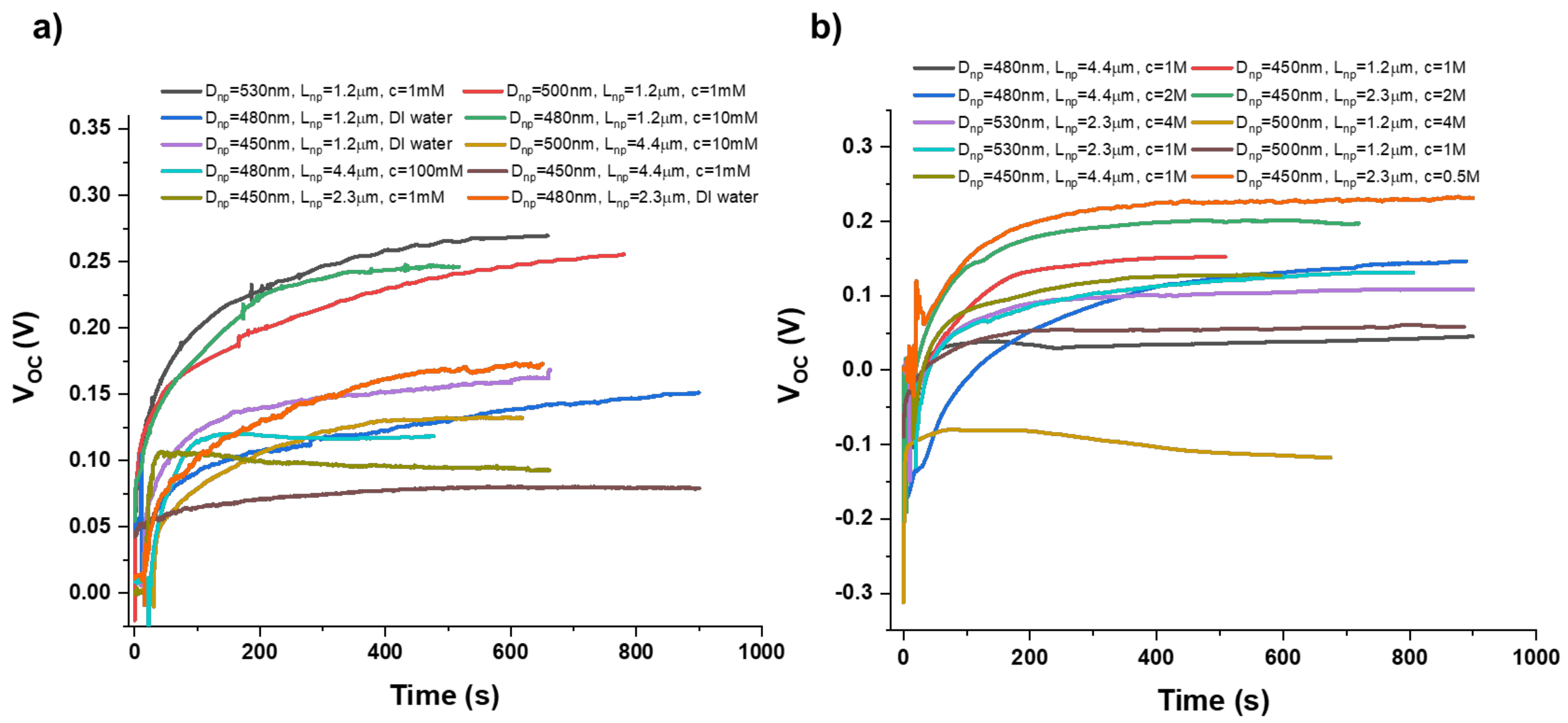

**Figure S3: a)** Open circuit voltage with time at low concentrations showing that the initial voltage is zero and stabilizes to a steady positive value. The different curve corresponds to a set of parameters namely concentration of the electrolyte, length, and diameter of the nanopillars. **b)** Open circuit voltage at higher concentrations, showing that the initial voltage is negative and its magnitude also increases with concentration. The final steady value measured was positive or negative depending on the set of parameters, but especially at the highest concentration. The electrolyte here is KCl in DI water.

## S4: Surface charge modelling

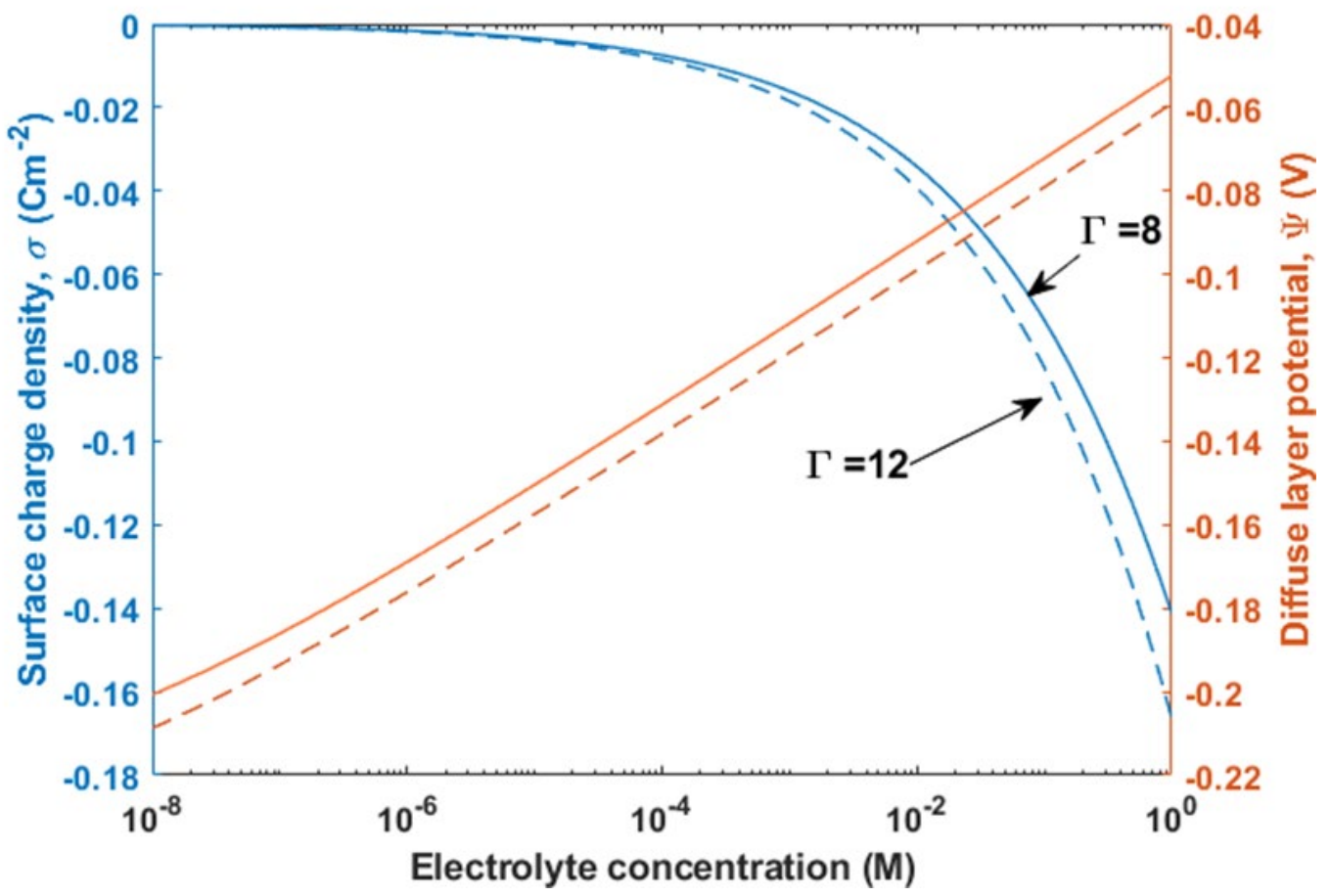

**Figure S4:** Surface charge density and diffuse layer potential for two different values of the parameter gamma. The chemical equilibrium and Grahame equation were solved for a particular value of Gamma as a function of electrolyte concentration that gives a unique surface charge density and diffuse layer potential value.[1]

$$\Psi_d(\sigma) = \frac{k_B T}{e}\left[ln\left(\frac{-\sigma}{\sigma + e\Gamma}\right) - \ln(10)(pH - pK)\right] - \frac{\sigma}{C} \qquad (1)$$

$$\sigma\,(\psi_d) = \frac{2\epsilon_0 \epsilon \kappa k_B T}{e}\left[\sinh\frac{e\psi_d}{2k_B T} + \frac{2}{\kappa r}\tanh\frac{e\psi_d}{4k_B T}\right] \qquad (2)$$

The equations were solved as a function of concentration at pH=7, pK=7.2, and C=0.29 $Fm^{-2}$, ε=78.5, T=300 K, and average radius of curvature, r=300 nm.

## S5: Open circuit voltage with concentration-independent surface charge

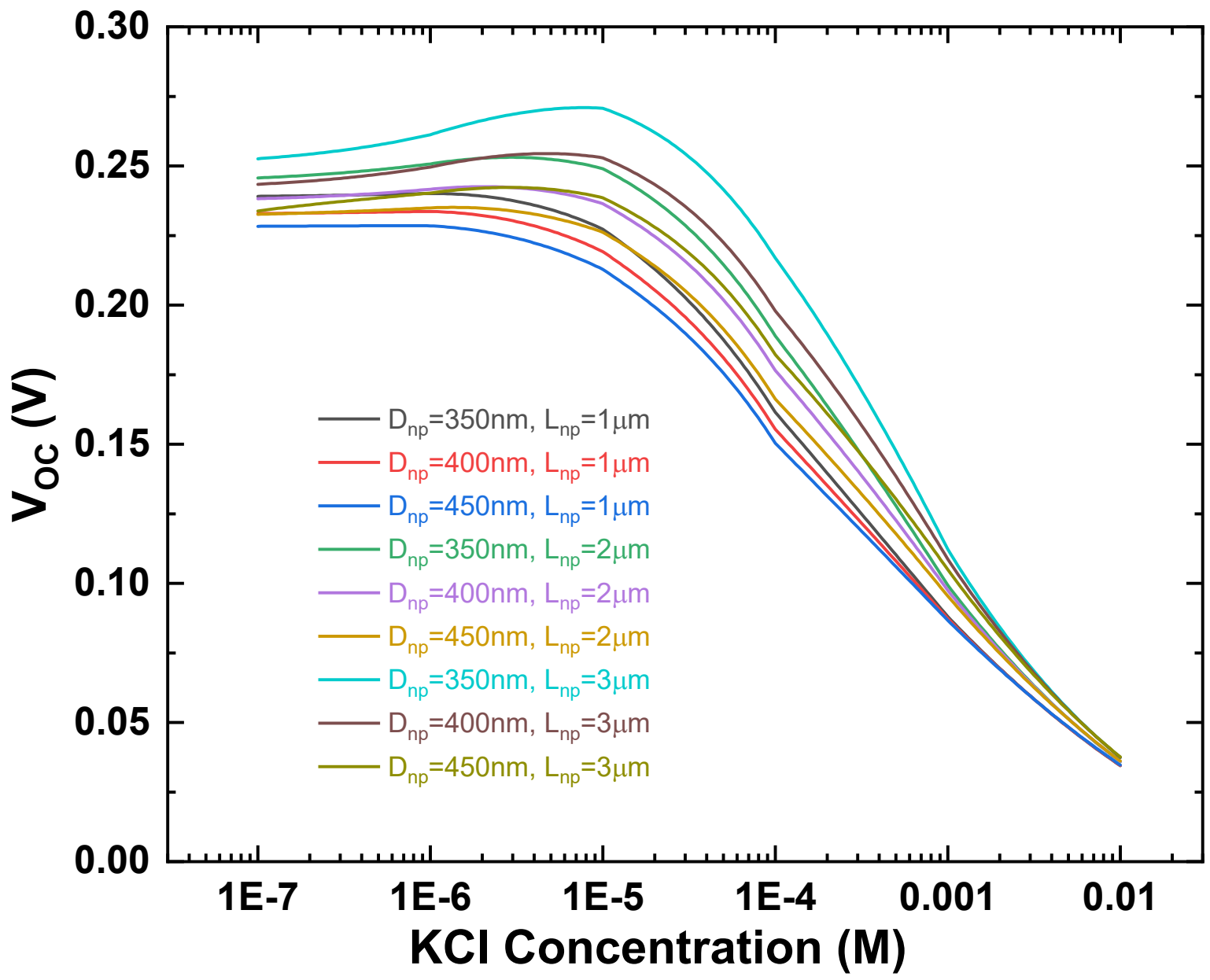

**Figure S5**: To elucidate the importance of concentration dependent surface charge, we performed simulation with a fixed surface charge of -10 mC/m$^2$. This result, which are monotonically decreasing with concentration, is in sharp contrast with the experimental trends. The monotonic decrease is due to increase in surface charge screening characterized by the Debye length, while the surface charge density is fixed.

**S6: Effect of pH**

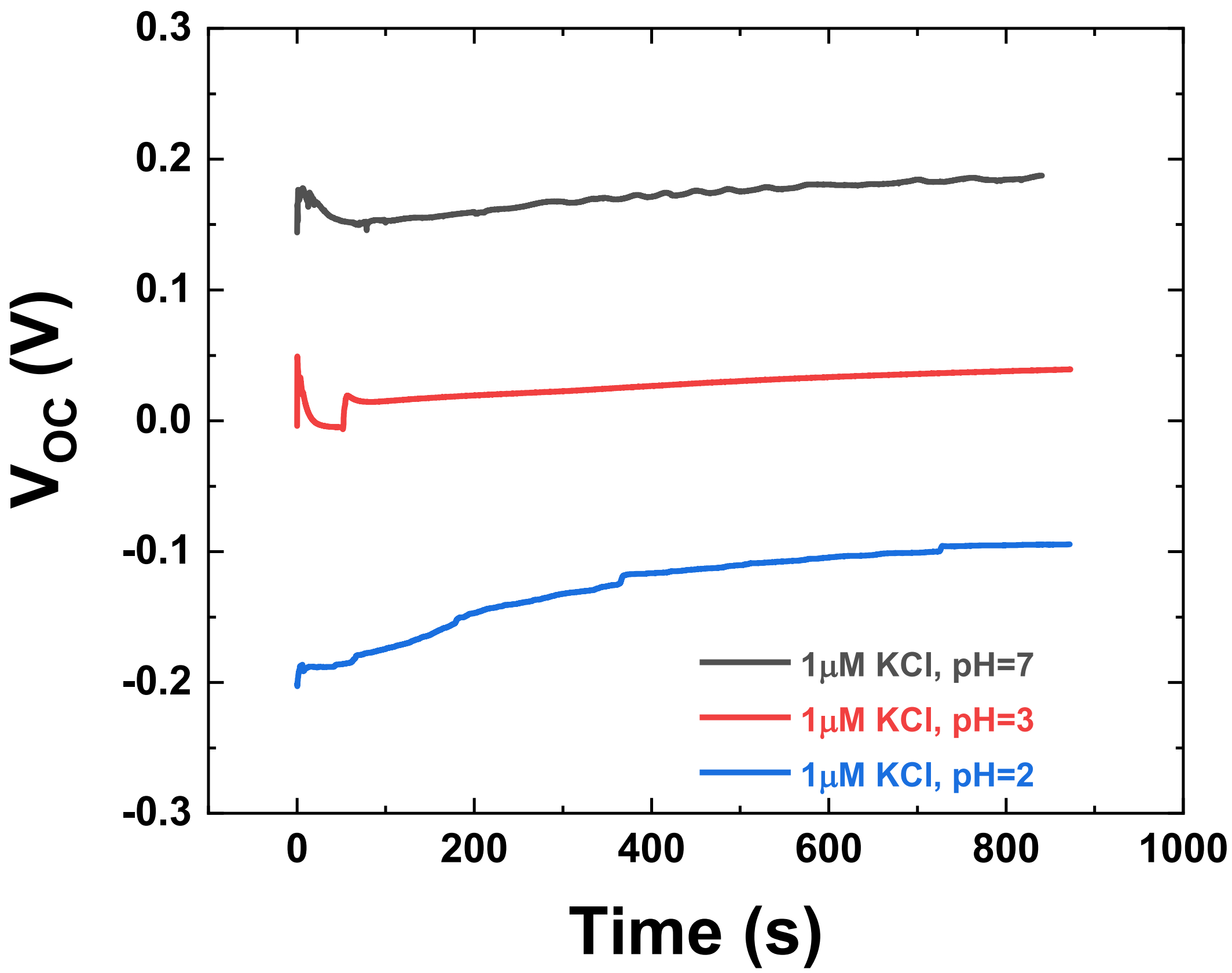

**Figure S6:** Time trace of the measured open circuit voltage for different pH at 1 μm KCl concentration. Different pH of the electrolyte was obtained by adding varying concentration of HCl to the original electrolyte (KCl in DI water). By changing the pH across the isoelectric point (pH=2-3) we observed both the change in sign and magnitude of open circuit voltage.

## S7: Electrochemical impedance spectroscopy

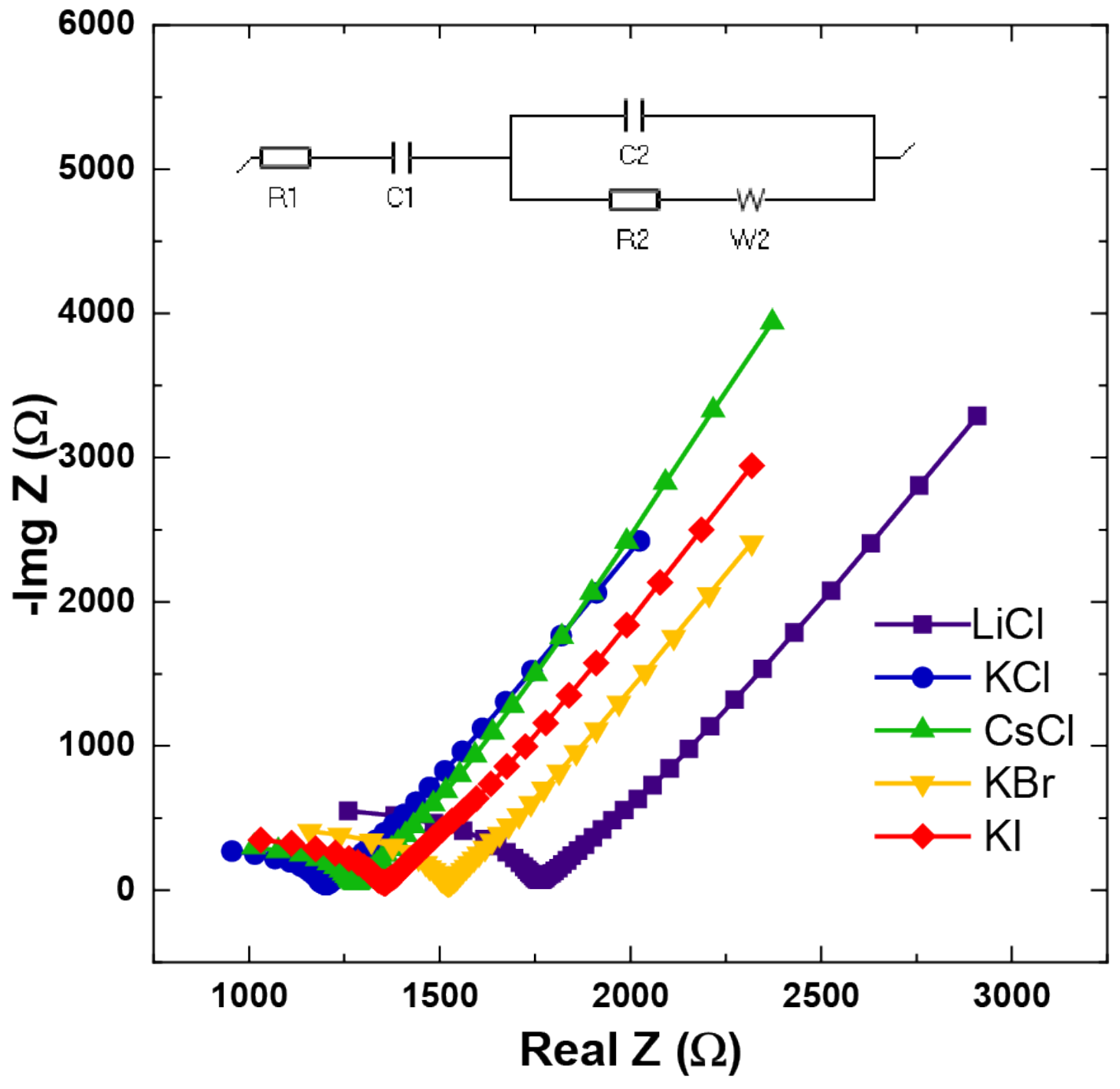

**Figure S7:** Nyquist plot for a single device and 1mM electrolyte concentration for different salt monovalent salts. The inset shows the equivalent electrical circuit used to fit the results and obtain the required circuit elements. The W2 is the Warburg impedance, which can be used to calculate the diffusivities.

| **Circuit parameters** | **LiCl** | **KCl** | **CsCl** | **KBr** | **KI** |
|---|---|---|---|---|---|
| **R1(Ω)** | 575.4 | 600.6 | 610.4 | 622.2 | 599.9 |
| **C1(F)** | 6.50E-07 | 5.88E-07 | 4.57E-07 | 8.66E-07 | 6.47E-07 |
| **C2(F)** | 1.00E-10 | 1.90E-10 | 1.70E-10 | 1.31E-10 | 1.60E-10 |
| **R2(Ω)** | 1132 | 565.2 | 624.9 | 858.3 | 715.4 |
| **W2=σ(Ωs$^{-0.5}$)** | 23602 | 16276 | 19051 | 16012 | 17993 |

The diffusion coefficient can be obtained from the following equation[2]:

$$\sigma = \frac{RT}{F^2 A\sqrt{2}D^{0.5}C} \qquad (3)$$

R = 8.314 J mol$^{-1}$ K$^{-1}$, T=300 K, $A_{S\text{-}L}$ =6.24 cm$^2$, C=1mM, F=96480 C mol$^{-1}$.

S8: IV characteristics and power

a)

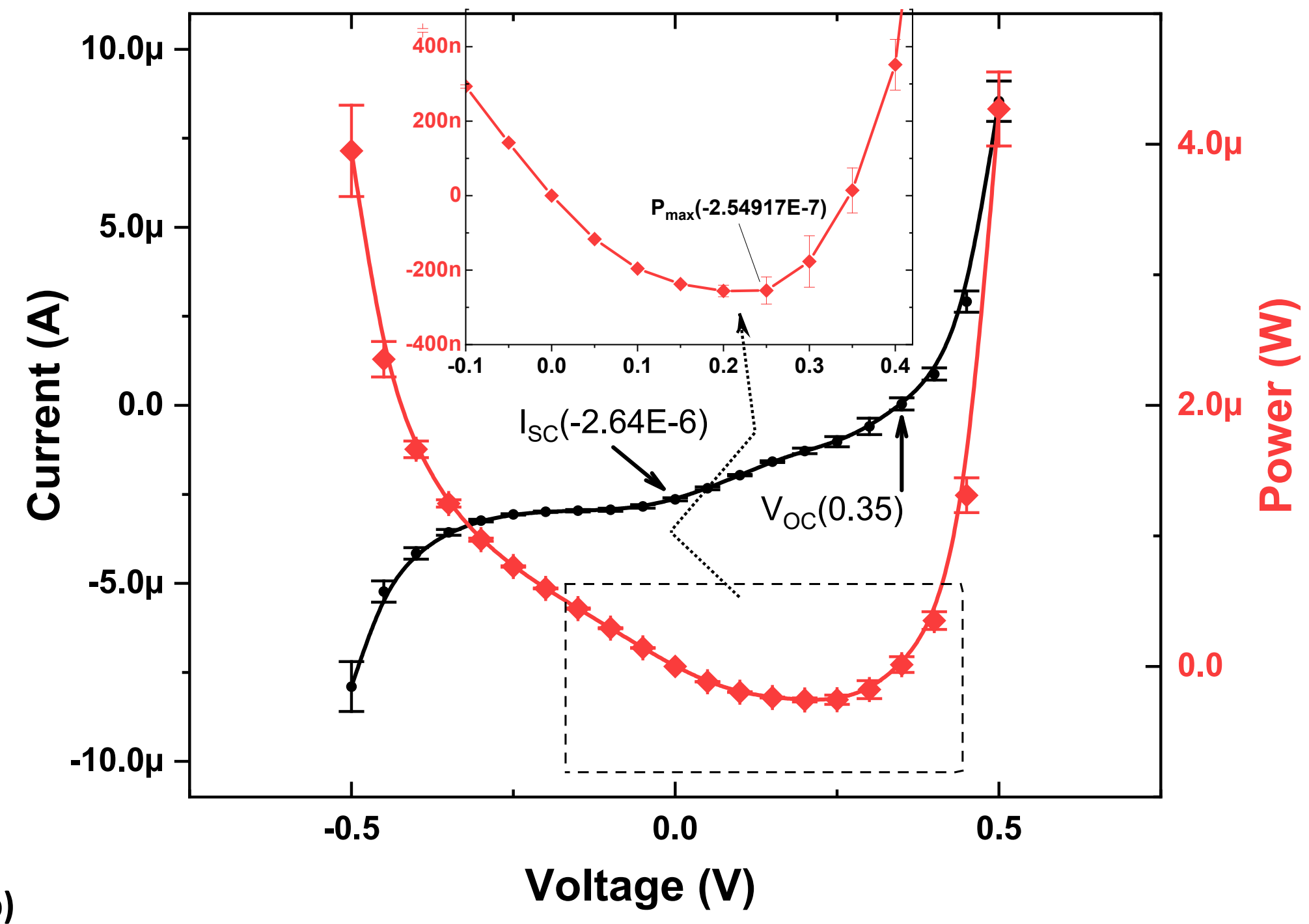

b)

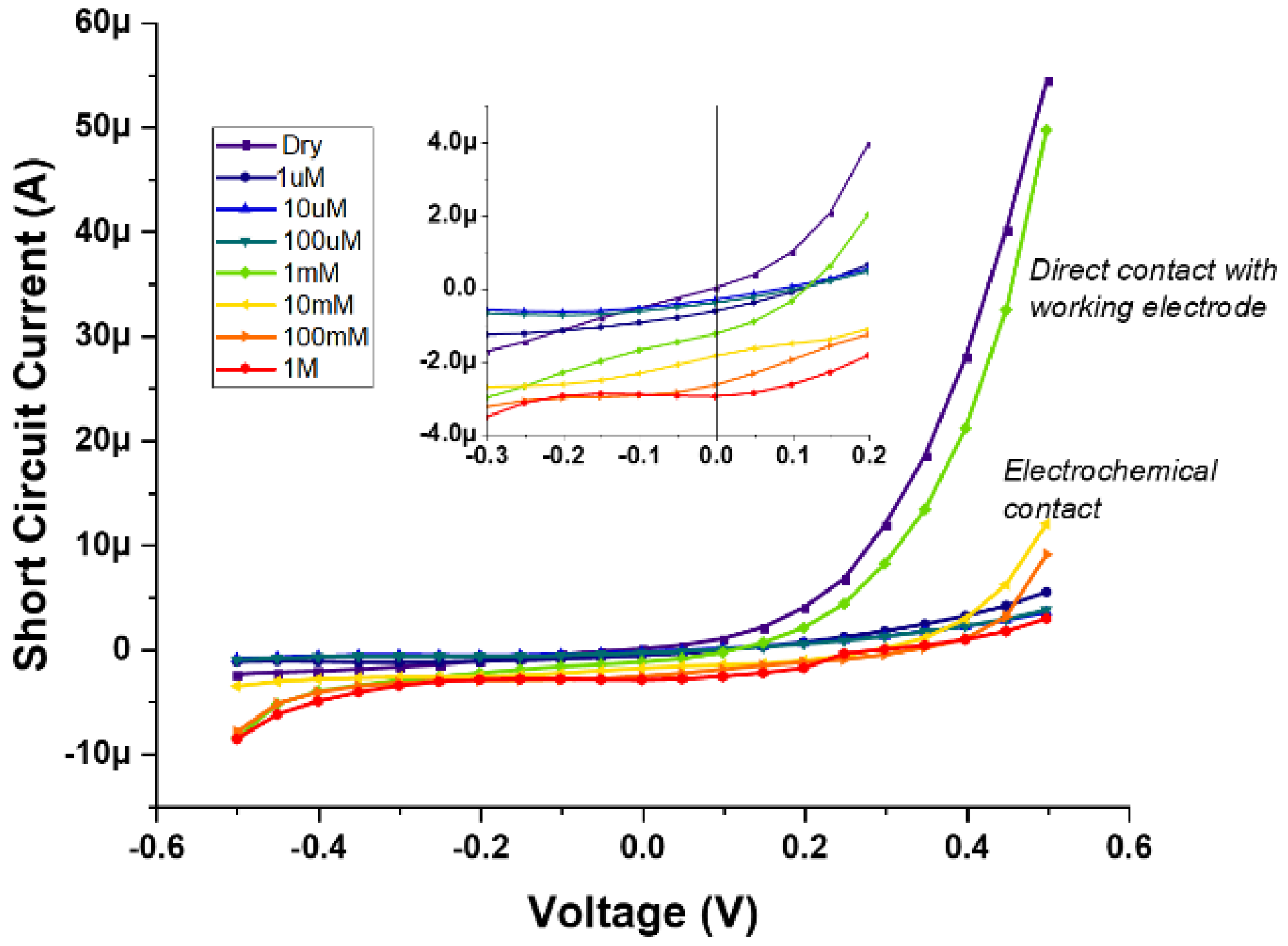

**Figure S8: a)** IV characteristics were obtained using Linear Sweep Voltammetry (LSV) by sweeping the voltage from -0.5V-0.5V. Using IV data, the maximum power output was determined. **b)** The IV characteristics for different concentration. Direct contact with electrochemical means that the Ag/AgCl is contacted directly with the silicon substrate, while in case of electrochemical contact, the Ag/AgCl electrode was put in the electrolyte 'film' on top of silicon substrate. In the Dry condition, the device behave as a diode with zero current at zero applied voltage. In presence of electrolyte, the curve shifts downwards having a negative short circuit current and positive open circuit voltage. For 1mM, we intentionally make a direct contact of Ag/AgCl wire with Si substrate, and we see a diode like behavior but with negative short current. The magnitude of current in this case is consistent with the increasing trend with concentration.

**S9: Initial and effective open circuit voltage measured at high concentration**

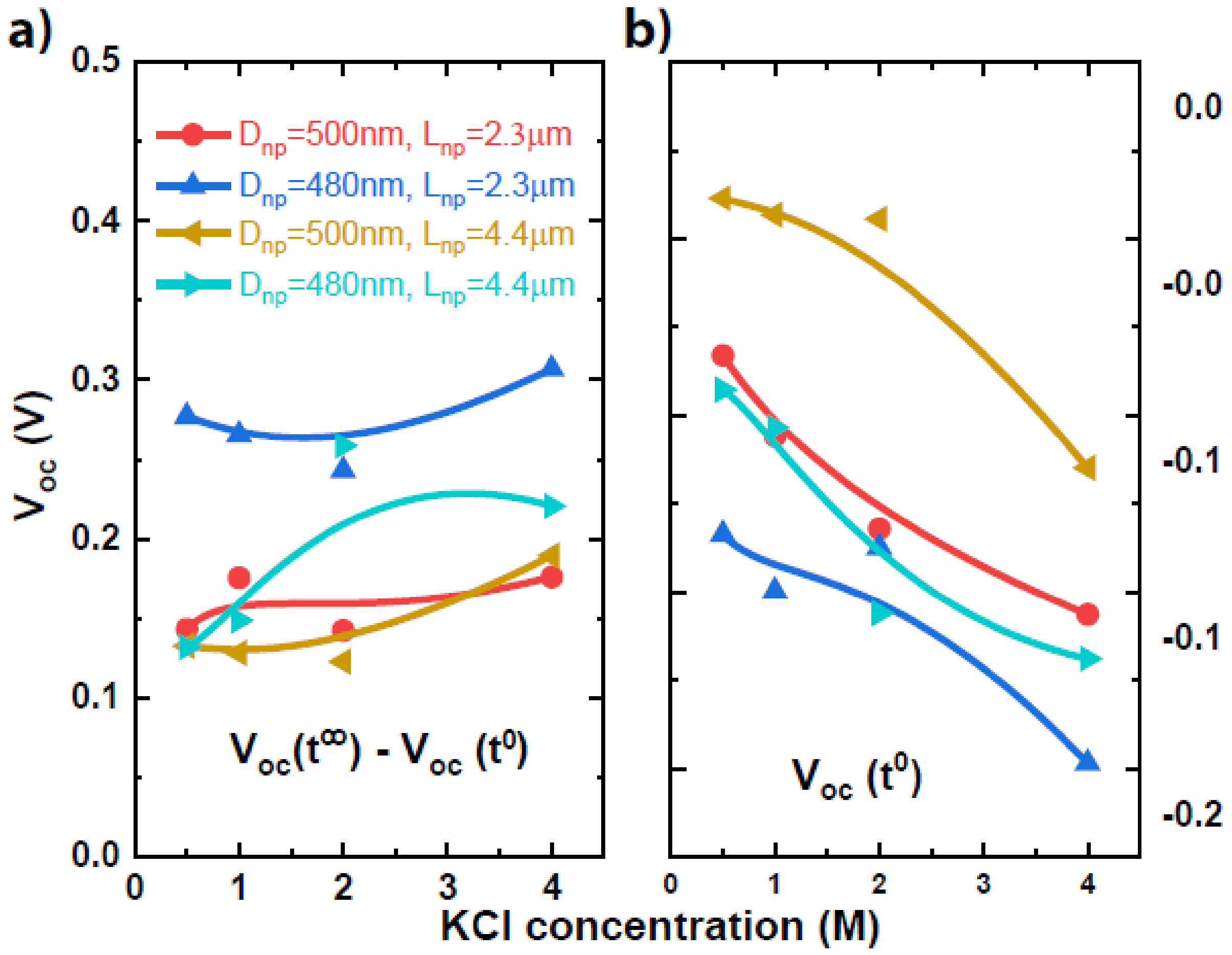

**Figure S9**: **a)** Effective open circuit voltage. **b)** initial open circuit voltage for different devices and range of KCl concentration from 0.5-4M.

$$V_{OC}^{eff} = V_{OC}^{\infty} - V_{OC}^{0} \qquad (4)$$

## S10: Silicon Nanopillars after long time testing

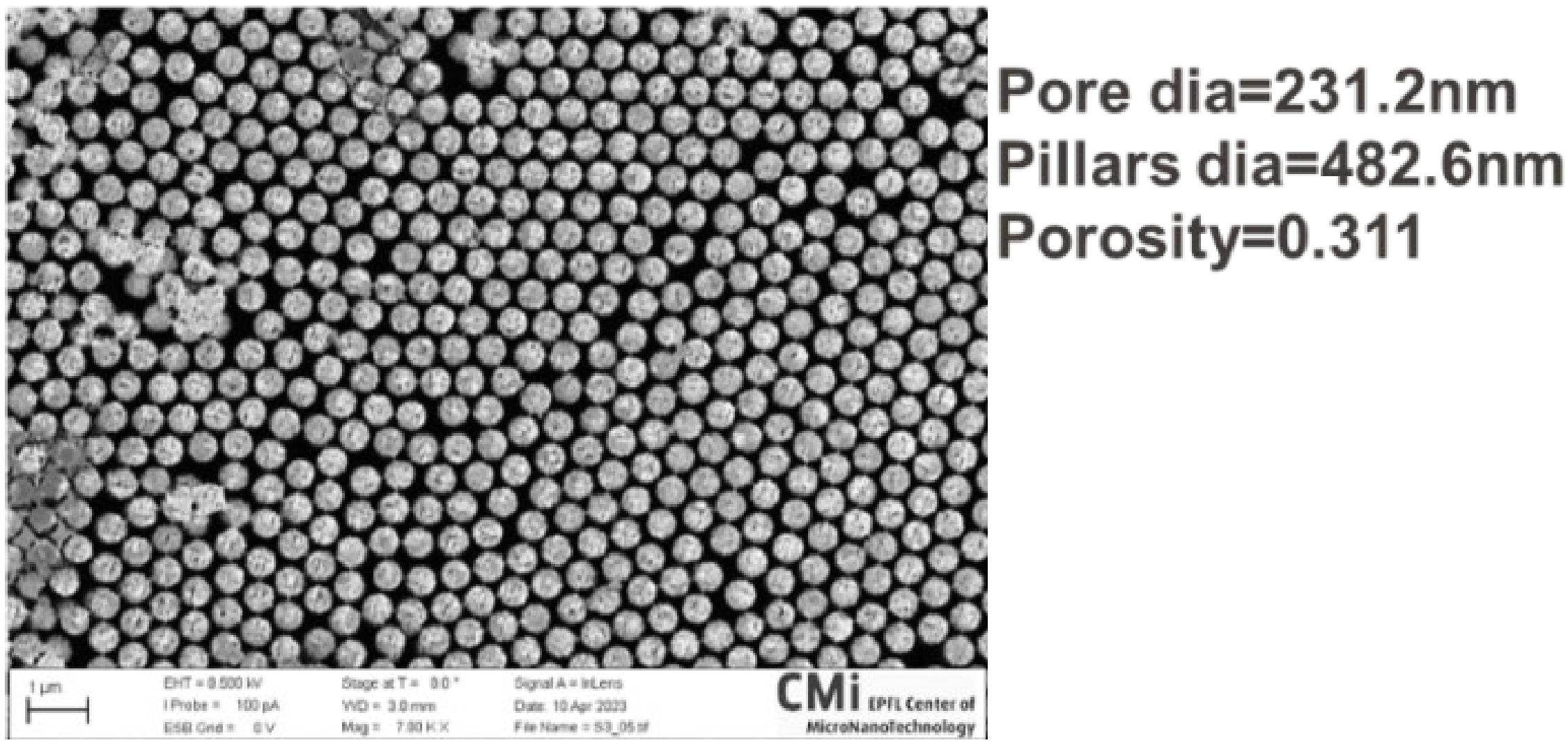

**Figure S10:** SEM image of silicon nanopillars after long time test under high concentrations (up to 4M) both open circuit test and current-voltage measurement. Two effects can affect long-term stability: i) salt crystallization, ii) reaction of silicon with the chloride ions. Yet, the image show a good stability of the studied structure.

## S11: Fabrication methodology

| Step | Process description | Cross-section after process |
|---|---|---|
| 01 | Substrate: *P-type Si*<br><br>*Self-assembly of Polystyrene(PS) particles (600nm)*<br><br>Thickness: *monolayer assembly* | 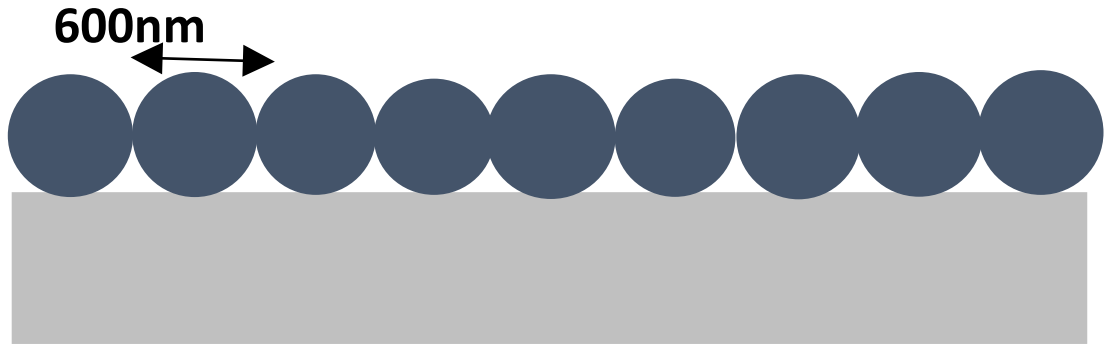 |
| 02 | *Oxygen Plasma etching*<br><br>Oxygen flow rate:5-10 ml/min<br><br>Power:50W<br><br>Time:5-15mins | 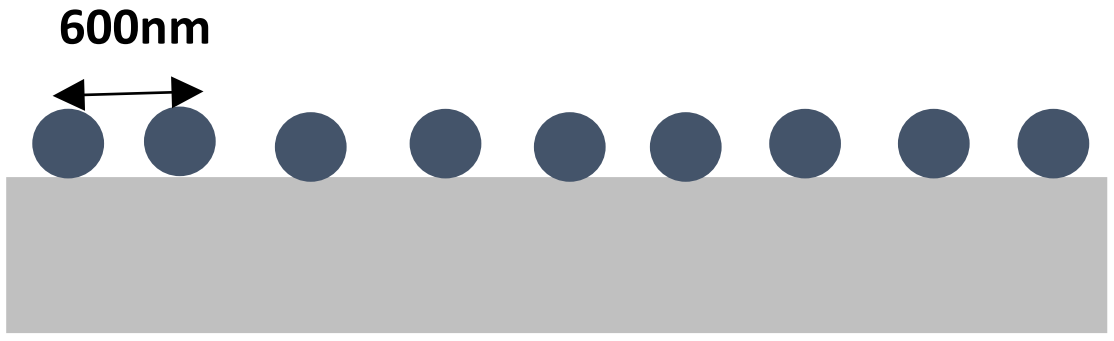<br>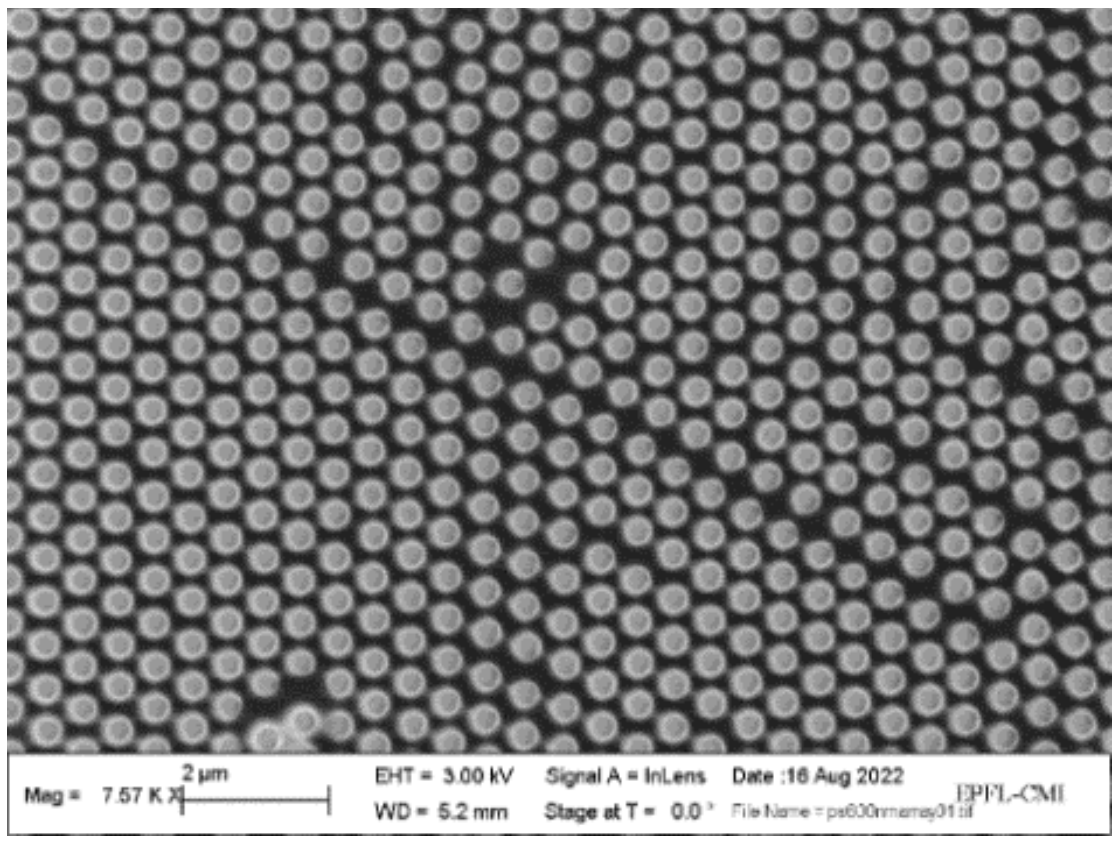 |
| 03 | *Metal Evaporation*<br><br>Material: *Ti*<br><br>Thickness:5nm | 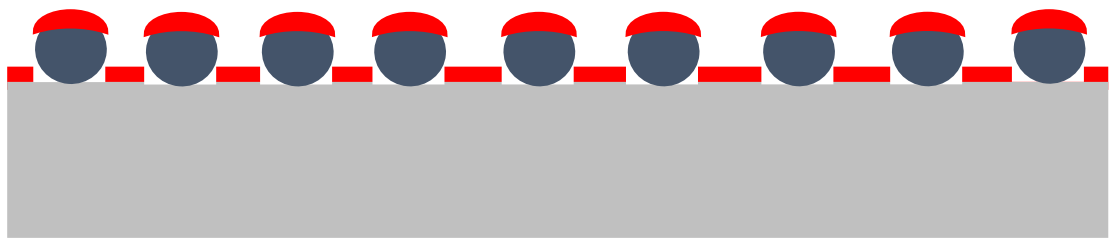 |
| 04 | ***Metal sputtering***<br>**Material :** ***Gold***<br>**Thickness :** ***25nm*** | 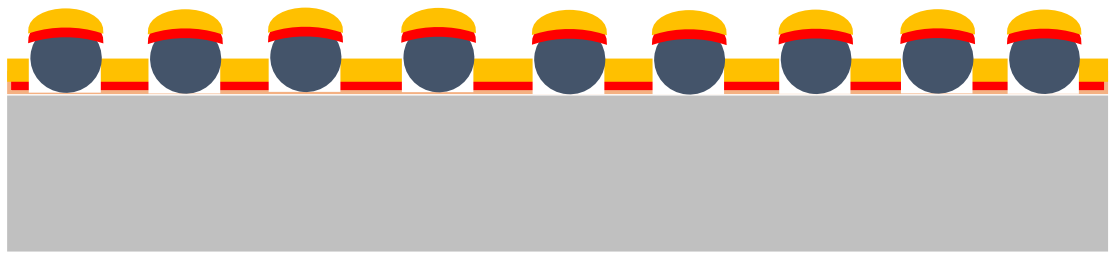 |
| 05 | ***Removal of PS particle,***<br>***solvent stripping***<br>***Solvent: Toulene*** | 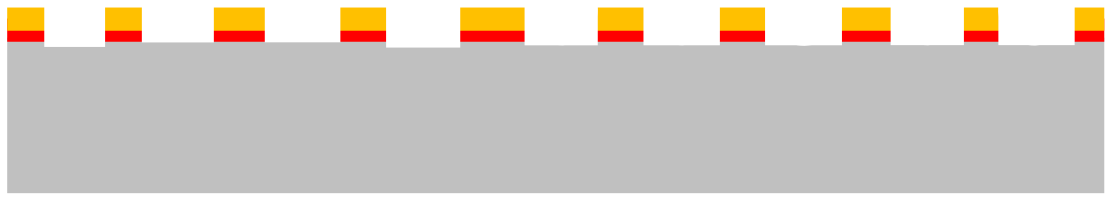<br>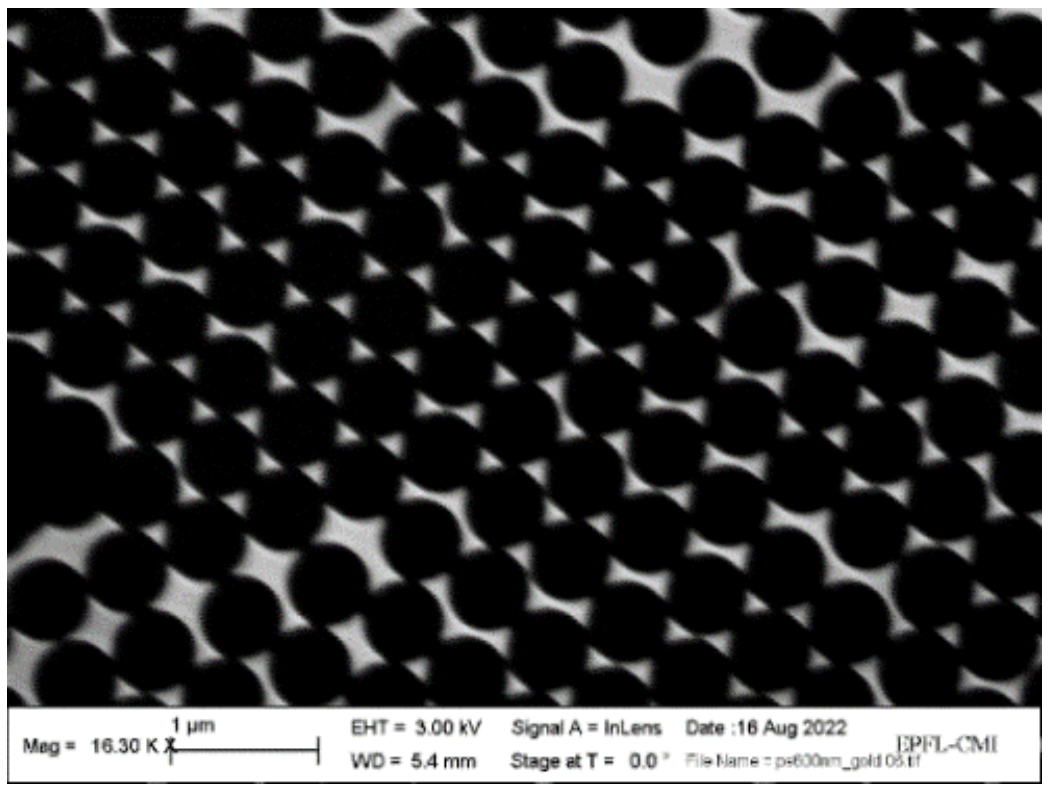<br> |
| 06 | ***Metal assisted chemical***<br>***etching of silicon***<br>**Concentration : HF = 10%,**<br>**$H_2O_2$ = 2%**<br>**Time : 5-20mins** | 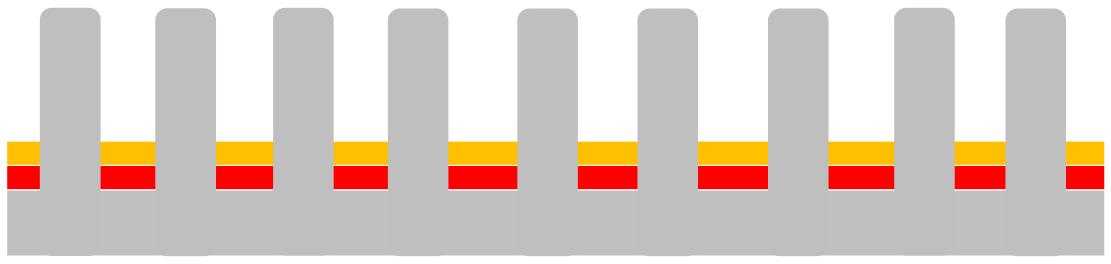 |
| 07 | ***Metal removal using aqueous***<br>***KI/$I_2$ solution(10% KI, 5% $I_2$)***<br>***Time: 1-2mins*** | 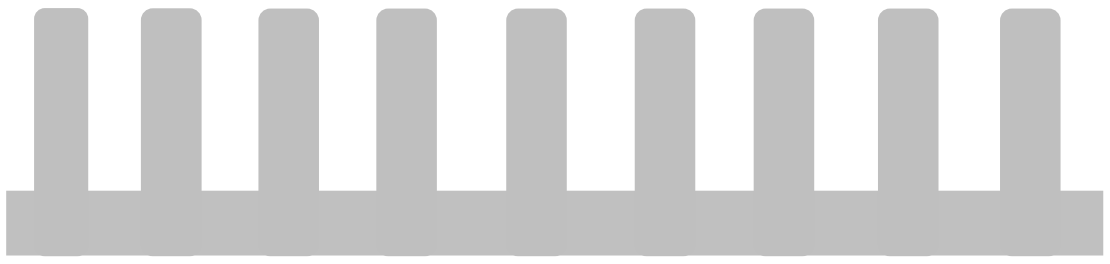 |
| | | 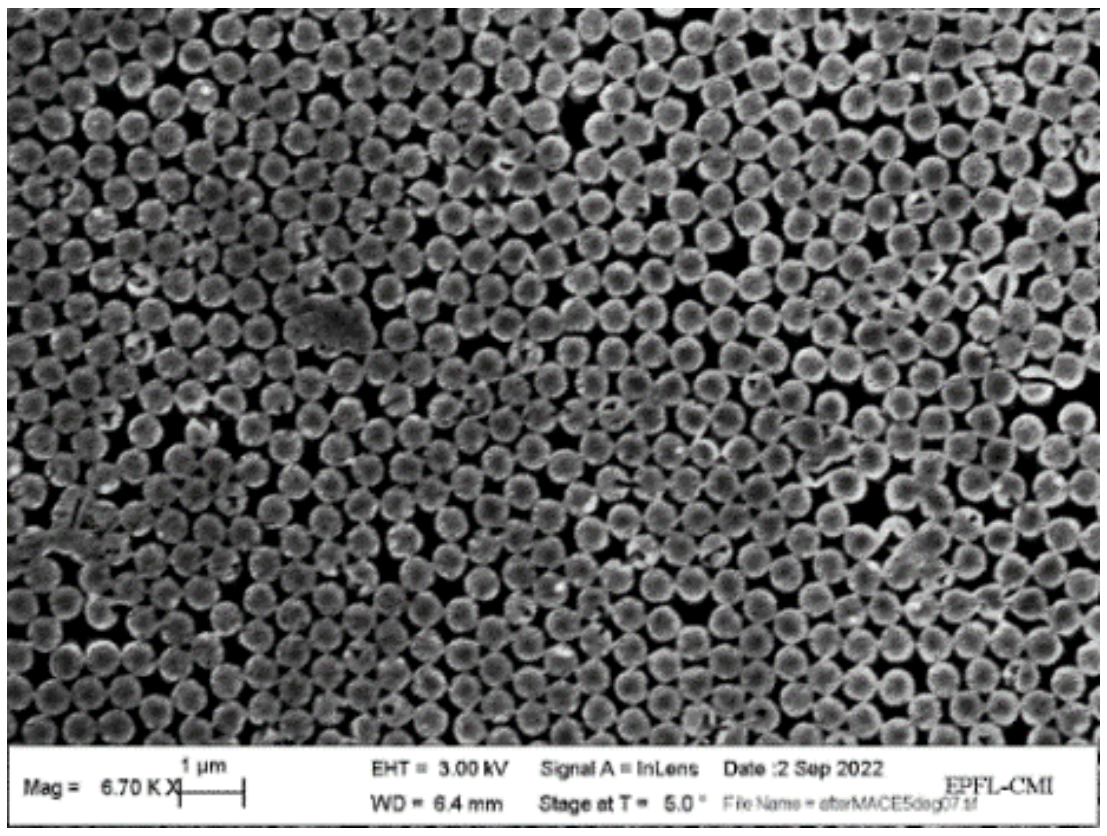 |

**Figure S11:** Fabrication of the Silicon NP using a combination of Colloidal lithography and metal assisted chemical etching[3]. The SEM image showing the sample after each step. **a)** Monolayer assembly of Polystyrene nanosphere after step 2. **b)** Gold Nano mesh after step 5. **c)** Silicon nanopillars after MACE in step 6.

## S12: Open circuit voltage measurement with different top electrode and planar silicon

**a)**

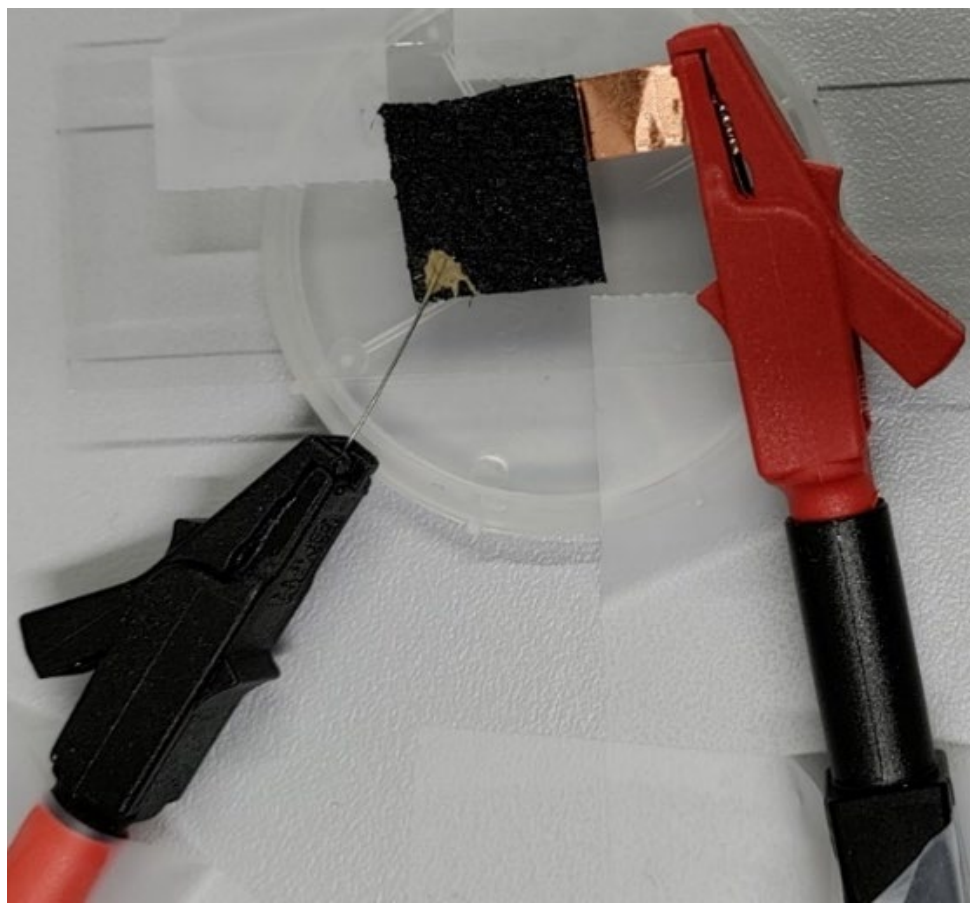

**b)**

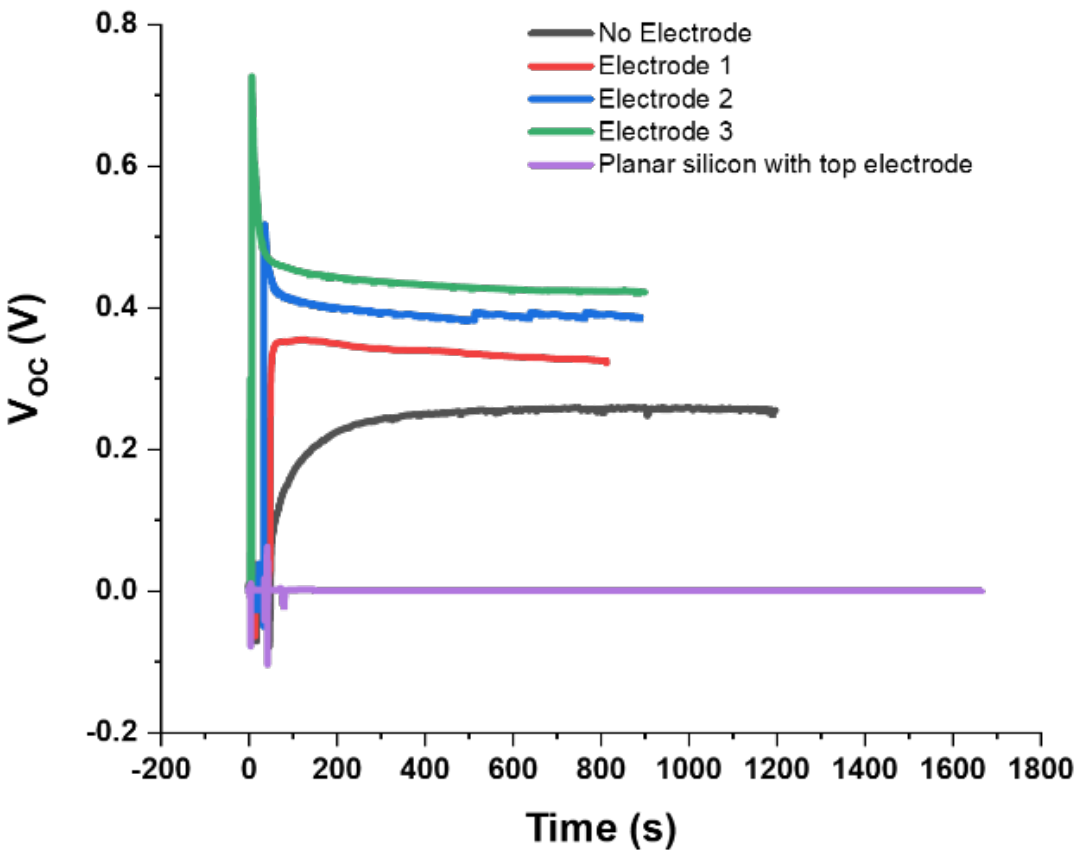

**Figure S12: a)** Measurement of open circuit voltage with a top porous electrode. **b)** The measured voltage measured in a single Silicon Nanopillars device with and without top porous electrode. Three different electrode were used that varied and sheet resistance and hydrophilicity and therefore different open circuit was obtained with the same device[4]. The voltage was also measured with top electrode on a planar silicon and it showed almost zero open circuit voltage.

**S13: Simulation details**

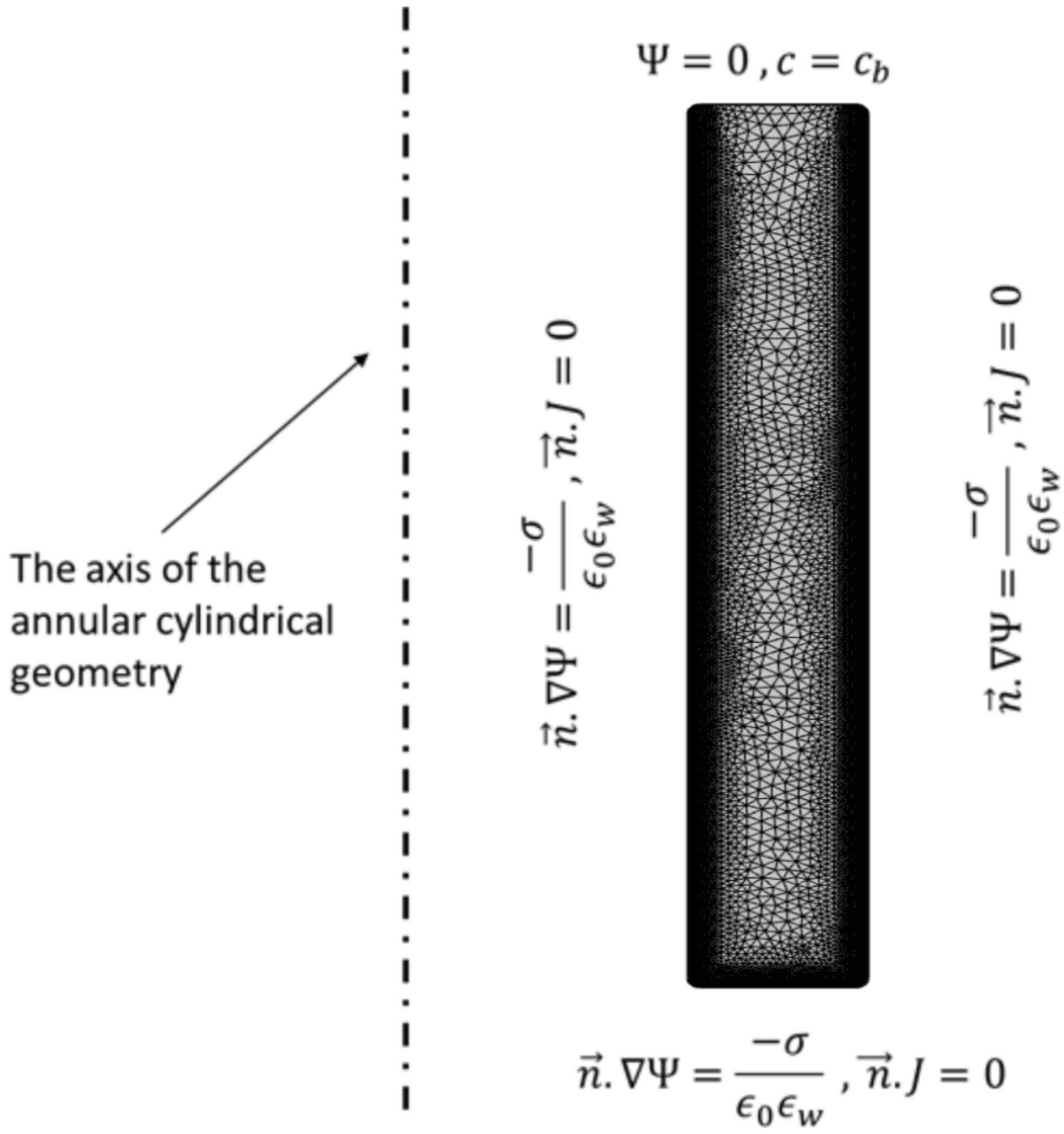

**Figure S13:** Geometry of the simulation, with the relevant boundary conditions, and meshing.

The simulation was performed using Comsol 6.0 using the transport of dilute species, and electrostatics packages that were coupled. The simulation was performed in 2D-axis symmetric domain as shown in the figure above. The normal flux of ions at the wall was set to zero, while at the top a concentration boundary condition was used. A concentration dependent surface charge was used as the boundary conditions at the solid surface as described in S4, while potential was set equal to zero at the top surface. The computed potential difference between top and bottom is equal to $V_{OC}$. The meshing was finer near the wall due to large gradients, while a slightly coarser mesh was used near the bulk to optimize the simulation time.

## S14: Ionic concentration distribution for two representative electrolyte concentrations

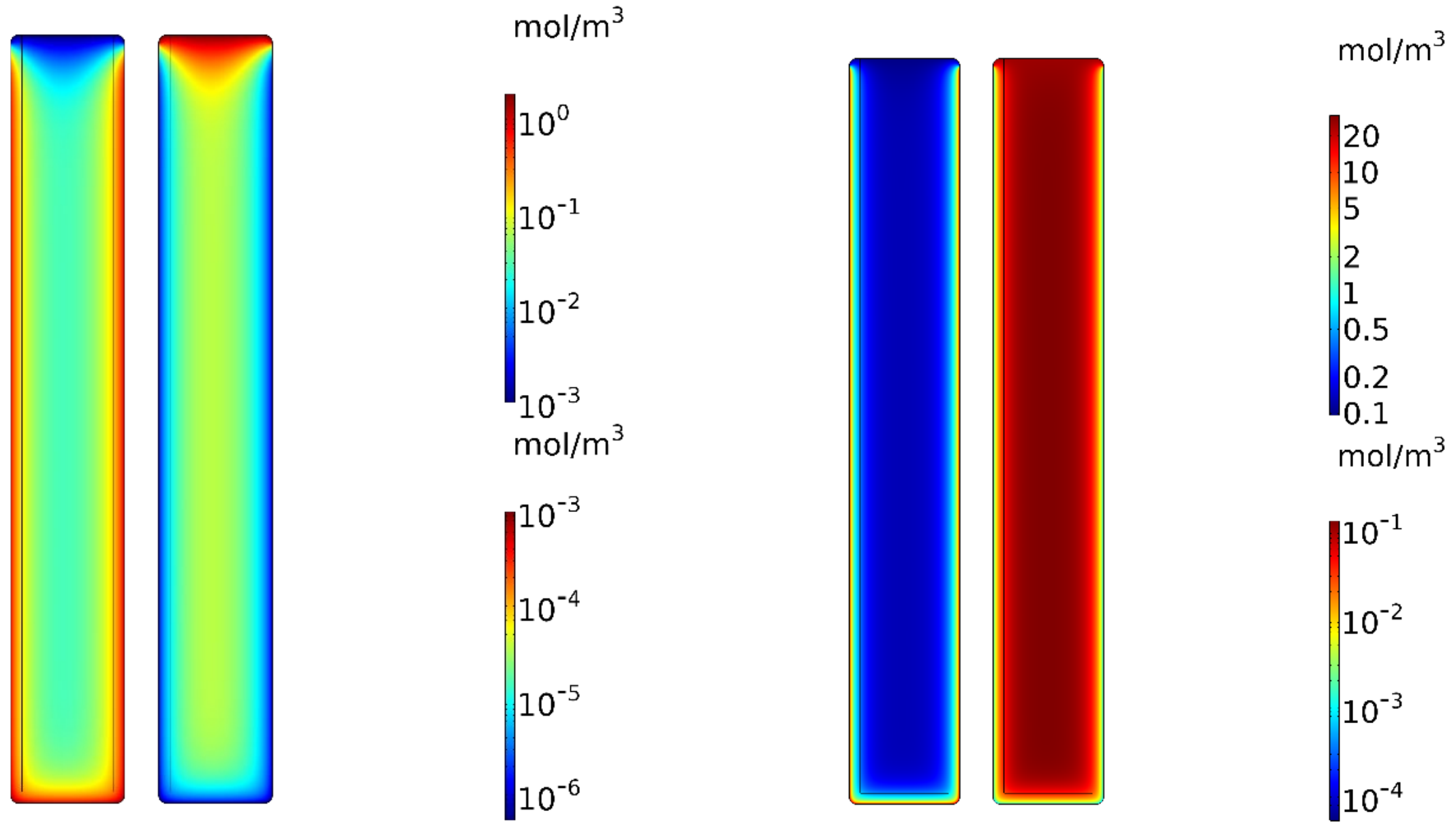

**Figure S14**: Cation and anion distribution for a negatively charged surface for a bulk electrolyte concentration of **1uM (left)** and **100uM (right)** KCl under purely static condition (with no convection). Concentration of cations close is to walls is much higher than in the bulk, while it is opposite for anions. For high bulk concentration, the electrical potential decays much faster due to small Debye-length, and therefore the concentration of ions is more uniformly distributed (lower concentration gradient) throughout the channel. For low bulk concentration, the electrical potential decays much slowly away from the surface, and therefore the ions concentrations have larger concentration gradients.

## References

1. Behrens, S. H. & Grier, D. G. The charge of glass and silica surfaces. *The Journal of Chemical Physics* **115**, 6716–6721 (2001).

2. Shi, M., Chen, Z. & Sun, J. Determination of chloride diffusivity in concrete by AC impedance spectroscopy. *Cement and Concrete Research* **29**, 1111–1115 (1999).

3. Wendisch, F. J., Rey, M., Vogel, N. & Bourret, G. R. Large-Scale Synthesis of Highly Uniform Silicon Nanowire Arrays Using Metal-Assisted Chemical Etching. *Chem. Mater.* **32**, 9425–9434 (2020).

4. Shao, B. *et al.* Bioinspired Hierarchical Nanofabric Electrode for Silicon Hydrovoltaic Device with Record Power Output. *ACS Nano* **15**, 7472–7481 (2021).